\pdfoutput=1

\documentclass[aps,prl,twocolumn,superscriptaddress,floatfix,preprintnumbers,nofootinbib]{revtex4-2}

\usepackage{amsmath, float, graphicx,amssymb,multirow,pgfplots,pgfplotstable}
\usepackage{subfig, float}
\usepackage{tikz,hyperref}
\usepackage{ulem}
\usepackage{comment}
\usepackage{url}
\usepackage{graphicx}
\usepackage{lipsum}

\usetikzlibrary{shapes.geometric,arrows}

\begin{document}

\begin{flushright}
\end{flushright}

\title{Aspects of Sommerfeld Enhancement in the light of Halo gamma-ray excess}

\author{Yongsoo Jho}
\email{1jys34@gmail.com}
\affiliation{Department of Physics and IPAP, Yonsei University, Seoul 03722, Republic of Korea}

\author{Jeonghwan Park}
\email{jnghwn.park@yonsei.ac.kr}
\affiliation{Department of Physics and IPAP, Yonsei University, Seoul 03722, Republic of Korea}

\author{Min Gi Park}
\email{mgpark@yonsei.ac.kr}
\affiliation{Department of Physics and IPAP, Yonsei University, Seoul 03722, Republic of Korea}

\author{Seong Chan Park}
\email{sc.park@yonsei.ac.kr}
\affiliation{Department of Physics and IPAP, Yonsei University, Seoul 03722, Republic of Korea}
\affiliation{School of Physics, Korea Institute for Advanced Study, Seoul, 02455, Republic of Korea}

\begin{abstract}
We examine Sommerfeld enhancement in dark matter annihilation as a potential origin of the halo‑like gamma‑ray excess near $E_\gamma\simeq 20$ GeV reported by Totani. A minimal model with a light CP‑even scalar mediator naturally produces a velocity‑dependent annihilation cross section consistent with thermal freeze‑out, the Milky Way excess, and limits from dwarf spheroidal galaxies. 
\end{abstract}

\maketitle


\newcommand{\PRE}[1]{{#1}} 
\newcommand{\ul}{\underline}
\newcommand{\del}{\partial}
\newcommand{\nbox}{{\,\lower0.9pt\vbox{\hrule \hbox{\vrule height 0.2 cm
\hskip 0.2 cm \vrule height 0.2 cm}\hrule}\,}}

\newcommand{\postscript}[2]{\setlength{\epsfxsize}{#2\hsize}
   \centerline{\epsfbox{#1}}}
\newcommand{\gweak}{g_{\text{weak}}}
\newcommand{\mweak}{m_{\text{weak}}}
\newcommand{\mplanck}{M_{\text{Pl}}}
\newcommand{\mstar}{M_{*}}
\newcommand{\sigmaan}{\sigma_{\text{an}}}
\newcommand{\sigmatot}{\sigma_{\text{tot}}}
\newcommand{\sigmaSI}{\sigma_{\rm SI}}
\newcommand{\sigmaSD}{\sigma_{\rm SD}}
\newcommand{\OmegaM}{\Omega_{\text{M}}}
\newcommand{\OmegaDM}{\Omega_{\text{DM}}}
\newcommand{\ipb}{\text{pb}^{-1}}
\newcommand{\ifb}{\text{fb}^{-1}}
\newcommand{\iab}{\text{ab}^{-1}}
\newcommand{\ev}{\text{eV}}
\newcommand{\kev}{\text{keV}}
\newcommand{\mev}{\text{MeV}}
\newcommand{\gev}{\text{GeV}}
\newcommand{\tev}{\text{TeV}}
\newcommand{\pb}{\text{pb}}
\newcommand{\mb}{\text{mb}}
\newcommand{\cm}{\text{cm}}
\newcommand{\m}{\text{m}}
\newcommand{\km}{\text{km}}
\newcommand{\kg}{\text{kg}}
\newcommand{\g}{\text{g}}
\newcommand{\s}{\text{s}}
\newcommand{\yr}{\text{yr}}
\newcommand{\Mpc}{\text{Mpc}}
\newcommand{\etal}{{\em et al.}}
\newcommand{\eg}{{\em e.g.}}
\newcommand{\ie}{{\em i.e.}}
\newcommand{\ibid}{{\em ibid.}}
\newcommand{\Eqref}[1]{Equation~(\ref{#1})}
\newcommand{\secref}[1]{Sec.~\ref{sec:#1}}
\newcommand{\secsref}[2]{Secs.~\ref{sec:#1} and \ref{sec:#2}}
\newcommand{\Secref}[1]{Section~\ref{sec:#1}}
\newcommand{\appref}[1]{App.~\ref{sec:#1}}
\newcommand{\figref}[1]{Fig.~\ref{fig:#1}}
\newcommand{\figsref}[2]{Figs.~\ref{fig:#1} and \ref{fig:#2}}
\newcommand{\Figref}[1]{Figure~\ref{fig:#1}}
\newcommand{\tableref}[1]{Table~\ref{table:#1}}
\newcommand{\tablesref}[2]{Tables~\ref{table:#1} and \ref{table:#2}}
\newcommand{\Dsle}[1]{\slash\hskip -0.28 cm #1}
\newcommand{\met}{{\Dsle E_T}}
\newcommand{\mpt}{\not{\! p_T}}
\newcommand{\Dslp}[1]{\slash\hskip -0.23 cm #1}
\newcommand{\Dsl}[1]{\slash\hskip -0.20 cm #1}

\newcommand{\mB}{m_{B^1}}
\newcommand{\mq}{m_{q^1}}
\newcommand{\mf}{m_{f^1}}
\newcommand{\mKK}{m_{KK}}
\newcommand{\WIMP}{\text{WIMP}}
\newcommand{\SWIMP}{\text{SWIMP}}
\newcommand{\NLSP}{\text{NLSP}}
\newcommand{\LSP}{\text{LSP}}
\newcommand{\mWIMP}{m_{\WIMP}}
\newcommand{\mSWIMP}{m_{\SWIMP}}
\newcommand{\mNLSP}{m_{\NLSP}}
\newcommand{\mchi}{m_{\chi}}
\newcommand{\mgravitino}{m_{\gravitino}}
\newcommand{\mmed}{M_{\text{med}}}
\newcommand{\gravitino}{\tilde{G}}
\newcommand{\Bino}{\tilde{B}}
\newcommand{\photino}{\tilde{\gamma}}
\newcommand{\stau}{\tilde{\tau}}
\newcommand{\slepton}{\tilde{l}}
\newcommand{\snu}{\tilde{\nu}}
\newcommand{\squark}{\tilde{q}}
\newcommand{\mgaugino}{M_{1/2}}
\newcommand{\epsEM}{\varepsilon_{\text{EM}}}
\newcommand{\mmess}{M_{\text{mess}}}
\newcommand{\lmess}{\Lambda}
\newcommand{\nmess}{N_{\text{m}}}
\newcommand{\signmu}{\text{sign}(\mu)}
\newcommand{\Omegachi}{\Omega_{\chi}}
\newcommand{\lambdafs}{\lambda_{\text{FS}}}
\newcommand{\be}{\begin{equation}}
\newcommand{\ee}{\end{equation}}
\newcommand{\bea}{\begin{eqnarray}}
\newcommand{\eea}{\end{eqnarray}}
\newcommand{\beq}{\begin{equation}}
\newcommand{\eeq}{\end{equation}}
\newcommand{\beqn}{\begin{eqnarray}}
\newcommand{\eeqn}{\end{eqnarray}}
\newcommand{\baln}{\begin{align}}
\newcommand{\ealn}{\end{align}}
\newcommand{\lsim}{\lower.7ex\hbox{$\;\stackrel{\textstyle<}{\sim}\;$}}
\newcommand{\gsim}{\lower.7ex\hbox{$\;\stackrel{\textstyle>}{\sim}\;$}}

\newcommand{\ssection}[1]{{\em #1.\ }}
\newcommand{\rem}[1]{\textbf{#1}}

\def\ie{{\it i.e.}\/}
\def\eg{{\it e.g.}\/}
\def\etc{{\it etc}.\/}
\def\calN{{\cal N}}

\def\mptwo{{m_{\pi^0}^2}}
\def\mp{{m_{\pi^0}}}
\def\sqtsn{\sqrt{s_n}}
\def\sqtsn{\sqrt{s_n}}
\def\sqtsn{\sqrt{s_n}}
\def\sqts0{\sqrt{s_0}}
\def\Dsqts{\Delta(\sqrt{s})}
\def\Omegatot{\Omega_{\mathrm{tot}}}

\newcommand{\changed}[2]{{\protect\color{red}\sout{#1}}{\protect\color{blue}\uwave{#2}}}

\newcommand{\mgp}[1]{{\color{red}[MGP:#1]}}
\newcommand{\jhp}[1]{{\color{purple}[JHP:#1]}}
\newcommand{\scp}[1]{{\color{blue}[SCP:#1]}}


\section{Introduction}

Indirect searches for dark matter (DM) using cosmic gamma-ray observations have been extensively pursued by the Fermi-LAT collaboration~\cite{Roszkowski:2017nbc}, with complementary coverage at TeV scales provided by ground-based observatories such as HAWC, H.E.S.S., MAGIC, and VERITAS~\cite{Fermi-LAT:2025gei}. 
These observational efforts have identified several tentative anomalies across vastly different energy scales, which have historically catalyzed significant theoretical developments in DM model building: Notable examples include the 3.5~keV X-ray line~\cite{Bulbul:2014sua, Boyarsky:2014jta}, the 511~keV line from the Galactic center~\cite{Teegarden:2004ct, Knodlseder:2005yq}, and the 120--130~GeV gamma-ray excess~\cite{Weniger:2012tx, Bringmann:2012vr}.  The interpretation of such signals often necessitates non-standard DM scenarios, such as decaying DM, monochromatic line emission, or velocity-dependent annihilation cross sections~\cite{Huh:2007zw, Park:2012xq, Park:2014olo, Lee:2014xua, Park:2015gdo, Park:2015ysf, Ibarra:2013cra, Cline:2014vsa}.

Along with this context, Totani~\cite{Totani:2025fxx} recently reported a detailed analysis of fifteen years of Fermi-LAT data, identifying a halo-like gamma-ray excess at $E_\gamma \simeq 20~\text{GeV}$ in the region $|l| < 60^\circ$ and $10^\circ < |b| < 60^\circ$. 

Based on the spatial distribution and spectral characteristics of the signal, the author proposed a DM interpretation with a mass range of $m_{\chi} \approx 500\text{--}800~\text{GeV}$, where the excess is primarily attributed to DM annihilation into $b\bar{b}$ and/or $W^+W^-$ final states. After a rigorous consideration of various Galactic background models, the inferred annihilation cross section was found to be $\langle \sigma v \rangle_{\rm Halo} \simeq (5\text{--}10) \times 10^{-25}~\text{cm}^3/\text{s}$. Notably, this value is significantly larger than the standard thermal freeze-out (FO) expectation, $\langle \sigma v \rangle_{\rm FO} \sim 3 \times 10^{-26}~\text{cm}^3/\text{s}$.

Explaining such a substantial discrepancy while maintaining consistency with thermal relic dynamics remains a significant theoretical challenge within a purely perturbative framework~\cite{Murayama:2025ihg}. $p$-wave annihilating DM offers a partial resolution, as its phase-space averaged cross section is velocity-suppressed~\cite{Choquette:2016xsw}:
\begin{align}
\langle \sigma v\rangle_{p\text{-wave}} \simeq \frac{1}{2}A_\sigma v^2,
\end{align}
where $A_\sigma$ is a constant. However, the non-trivial velocity profiles across different environments—ranging from $v_{\rm FO} \sim 0.3$ at freeze-out and $v_{\rm Halo} \sim 10^{-3}$ in the Galactic halo to less than $5\times 10^{-5}$ in dwarf spheroidal galaxies (dSphs)~\cite{Walker:2007ju, Dehnen:2006cm}—require a more robust mechanism to bridge the gap between $\langle \sigma v \rangle_{\rm FO}$ and $\langle \sigma v \rangle_{\rm Halo}$. 

A natural solution arises from the Sommerfeld enhancement (SE) due to the presence of a light mediator. The SE is a non-perturbative phenomenon characterized by the amplification of the initial-state wave function in scattering or annihilation processes, typically mediated by the exchange of light messengers or long-range forces. Originally introduced in the 1930s for atomic physics~\cite{Sommerfeld:1931qaf}, the effect was later revisited in DM phenomenology to accurately treat the annihilation cross sections of WIMPs with masses near or above the TeV scale~\cite{Baer:1998pg, Hisano:2003ec, Hisano:2004ds, Hisano:2005ec, Cirelli:2007xd, Cirelli:2008pk, Arkani-Hamed:2008hhe, Pospelov:2008jd, Fox:2008kb, Iengo:2009ni, Feng:2010zp}. In this letter, we demonstrate how the velocity-dependent SE factor and the inherent $p$-wave suppression cooperatively provide a realistic boost to the cross section, satisfying both the halo excess and the relic density requirements.

The remainder is structured as follows: We first introduce a model of fermionic dark matter ($\chi$) coupled to the Standard Model (SM) via a light scalar mediator ($\varphi$). We then examine how the Sommerfeld enhancement  and \textit{resonant annihilation} near quasi-bound states efficiently modify the standard freeze-out dynamics. Subsequently, we present a quantitative evaluation of the velocity-dependent cross section and the resulting gamma-ray signatures explaining the $20~\text{GeV}$ halo excess. Supplementary details regarding our numerical and analytical treatment of the SE are provided in the Appendix.

\section{Model}

\begin{figure}[t]
\includegraphics[width=0.5\textwidth]{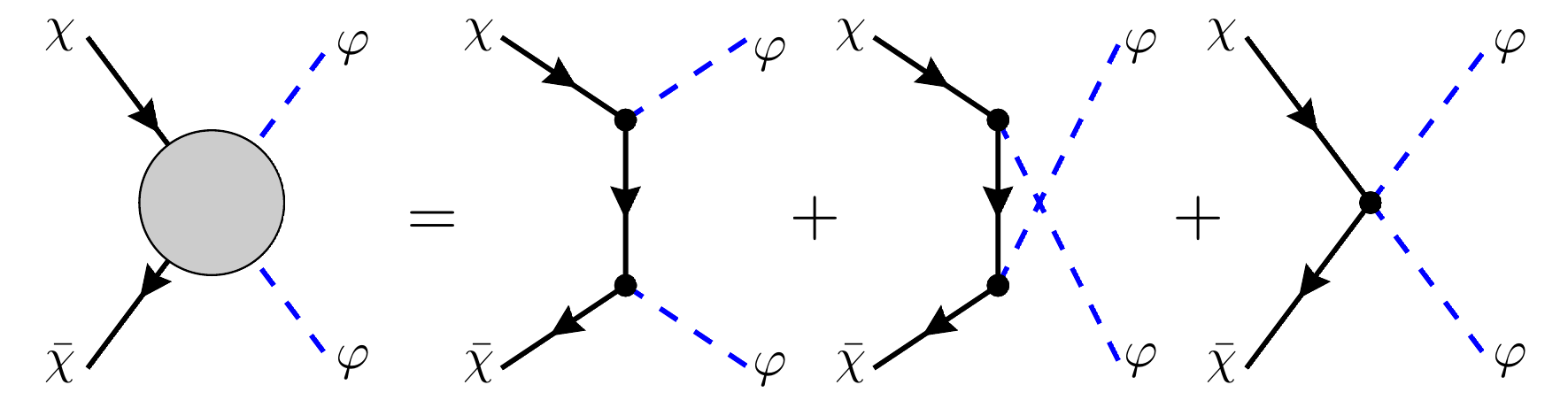}
\caption{The tree-level annihilation process $\chi \bar{\chi} \to \varphi \varphi$.}
\label{fig:hard_process}
\end{figure}

Let us consider an effective interaction between the Dirac fermionic dark matter (DM) $\chi$ and the light CP-even scalar $\varphi$ (mediator):
\begin{align}
\mathcal{L}_{\rm eff} &\supset  m_\chi e^{-\varphi/f_\varphi}\bar{\chi}\chi  
\label{eq:Yukawa}
\\
&= - \frac{m_\chi}{f_\varphi} \varphi \bar{\chi} \chi - \frac{m_\chi}{2 f_\varphi^2} \varphi^2 \bar{\chi} \chi +\cdots
\end{align}
where $f_\varphi =\langle \varphi \rangle$. Note that this dilaton-like interaction between DM and the scalar mediator $\varphi$ can appear, for instance, in spontaneously broken conformally invariant composite sectors \cite{Coleman:1973jx, Gildener:1976ih, Brivio:2017vri,  Bellazzini:2012vz, Bellazzini:2013fga, Chacko:2012sy, Holthausen:2013ota}. The presence of this interaction is crucial to specify angular momentum of the system in the annihilation process and ensures the $p$-wave dominance in the annihilation $\chi\bar{\chi} \to \varphi\varphi$ in Fig.~\ref{fig:hard_process}.  The cross section is given as
\begin{align}
\langle \sigma v \rangle_{\chi\chi\to \varphi\varphi} =bv^2 +{\cal O}(v^4),
\end{align}
where $s$-wave contribution is chirally forbidden due to the given interaction structure~\cite{Choquette:2016xsw}. For a typical WIMP with $m_\chi \sim \mathcal{O}(100)$ GeV, the observed relic density $\Omega_\chi h^2 \simeq 0.12$ requires $\langle \sigma v \rangle_{\rm FO} \simeq 3 \times 10^{-26} \, \text{cm}^3/\text{s}$ at $v_{\rm FO} \sim 0.3$. 

Following the analysis by Totani~\cite{Totani:2025fxx}, we consider the decay channel $\varphi \to b\bar{b}$, which leads to gamma-ray emission through the fragmentation and hadronization of bottom quarks, predominantly via neutral pion production. This scenario is naturally realized through a Higgs portal interaction, such as $\lambda_{\varphi H} \varphi^2 |H|^2$ or $\mu_{\varphi H} \varphi |H|^2$. These terms induce a mixing between the scalar mediator $\varphi$ and the Standard Model (SM) Higgs boson, characterized by a mixing angle $\theta$. The resulting effective coupling between the mediator and bottom quarks is given by:
\begin{align}
\mathcal{L} \supset y_b \sin\theta \, \varphi \bar{b}b,
\end{align}
where $y_b = \sqrt{2} m_b / v$ is the SM bottom-quark Yukawa coupling. In the following analysis, we assume a dominant branching fraction of $\text{Br}(\varphi \to b\bar{b}) \approx 100\%$. While alternative channels, such as $\varphi \to \tau^+\tau^-$, remain phenomenologically viable and could potentially alleviate the stringent constraints arising from cosmic-ray antiproton observations~\cite{AMS:2016oqu,Wang:2025zth}, we focus on the $b\bar{b}$ final state to maintain a direct comparison with the benchmark scenario presented in Ref.~\cite{Totani:2025fxx}.

\section{Sommerfeld Enhancement}

\begin{figure}[t]
\includegraphics[width=0.5\textwidth]{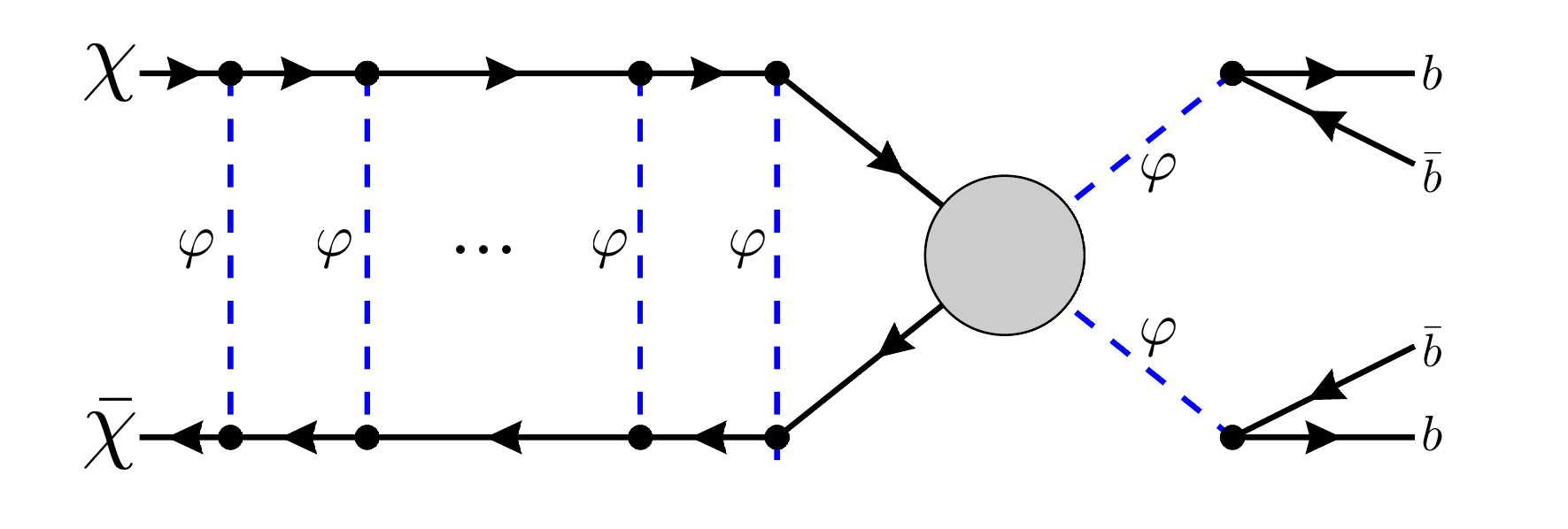}
\caption{DM annihilation in the presence of the Sommerfeld enhancement induced by long-range potential by $\varphi$. The gray blob indicates the hard process at short ranges shown in Fig.~\ref{fig:hard_process}. After that the production of a pair of scalar $\varphi$ in on-shell, subsequent decays provide gamma-ray signals.}
\label{fig:DM_annihilation}
\end{figure}

In the presence of a light mediator $\varphi$, the dark matter (DM) annihilation rate is non-perturbatively increased by the Sommerfeld effect, corresponding to the resummation of ladder diagrams (see Fig.~\ref{fig:DM_annihilation}). At low velocities $v \ll \alpha_{\rm eff} \ll 1$, this interaction is governed by the Yukawa potential $V(r) = -\alpha_{\rm eff} e^{-m_\varphi r}/r$, where the effective coupling is $\alpha_{\rm eff} \simeq m_\chi^2/(4\pi f_\varphi^2)$. We neglect subdominant $\mathcal{O}(1/f_\varphi^2 r^2)$ contact terms, as they do not contribute to the long-range dynamics driving the enhancement.

For a general partial wave with orbital angular momentum $l$, the enhancement factor $S_l$ is defined by the distortion of the wave function at the origin~\cite{Cassel:2009wt, Iengo:2009ni, Slatyer:2009vg}:
\begin{equation}
S_l = \left| \frac{(2l+1)!!}{|\mathbf{p}|^l \, l!} \left. \frac{\partial^l R_{El}(r)}{\partial r^l} \right|_{r=0} \right|^2,
\end{equation}
where $\mathbf{p}$ is the relative momentum and $R_{El}(r)$ is the radial solution to the Schrödinger equation. The total velocity-averaged cross section is the sum of these enhanced partial-wave contributions: $\langle \sigma v \rangle = \sum_l S_l \cdot \langle \sigma v \rangle_{(l)}^{\text{pert}}$, where $\langle \sigma v \rangle_{(l)}^{\text{pert}} \simeq a_l v^{2l}$ represents the short-range perturbative piece.

In the Coulomb limit ($m_\varphi \to 0$), the factor exhibits the asymptotic scaling $S_l(v) \propto (\alpha_{\rm eff}/v)^{2l+1}$~\cite{Cassel:2009wt}. For $p$-wave dominated annihilation ($l=1$), this yields a net scaling $\langle \sigma v \rangle \propto v^2 \times v^{-3} \sim v^{-1}$. This $1/v$ enhancement at galactic velocities ($v \sim 10^{-3}$) effectively compensates for the inherent $p$-wave suppression, providing the required boost to explain the Fermi-LAT halo excess while remaining consistent with the thermal relic abundance ($v \sim 0.3$).

The velocity dependence is further modified in the infrared by resonant scattering. Near the threshold of a zero-energy quasi-bound state, the enhancement saturates at a velocity $v_{\rm sat} \sim |\alpha_{\rm eff} - \alpha_{\rm crit}| \equiv \Delta \alpha$~\footnote{For a Yukawa potential, $p$-wave resonance occurs at $\alpha_{\rm crit} \approx 9.08 \, (m_\varphi / m_\chi)$. We solve the Schrödinger equation numerically for the exact Yukawa potential to determine the wave function and SE factor.}. As illustrated in Fig.~\ref{fig:xsec_velocity_dep}, we identify three distinct physical regimes:

\begin{figure}[h]
\includegraphics[width=0.45\textwidth]{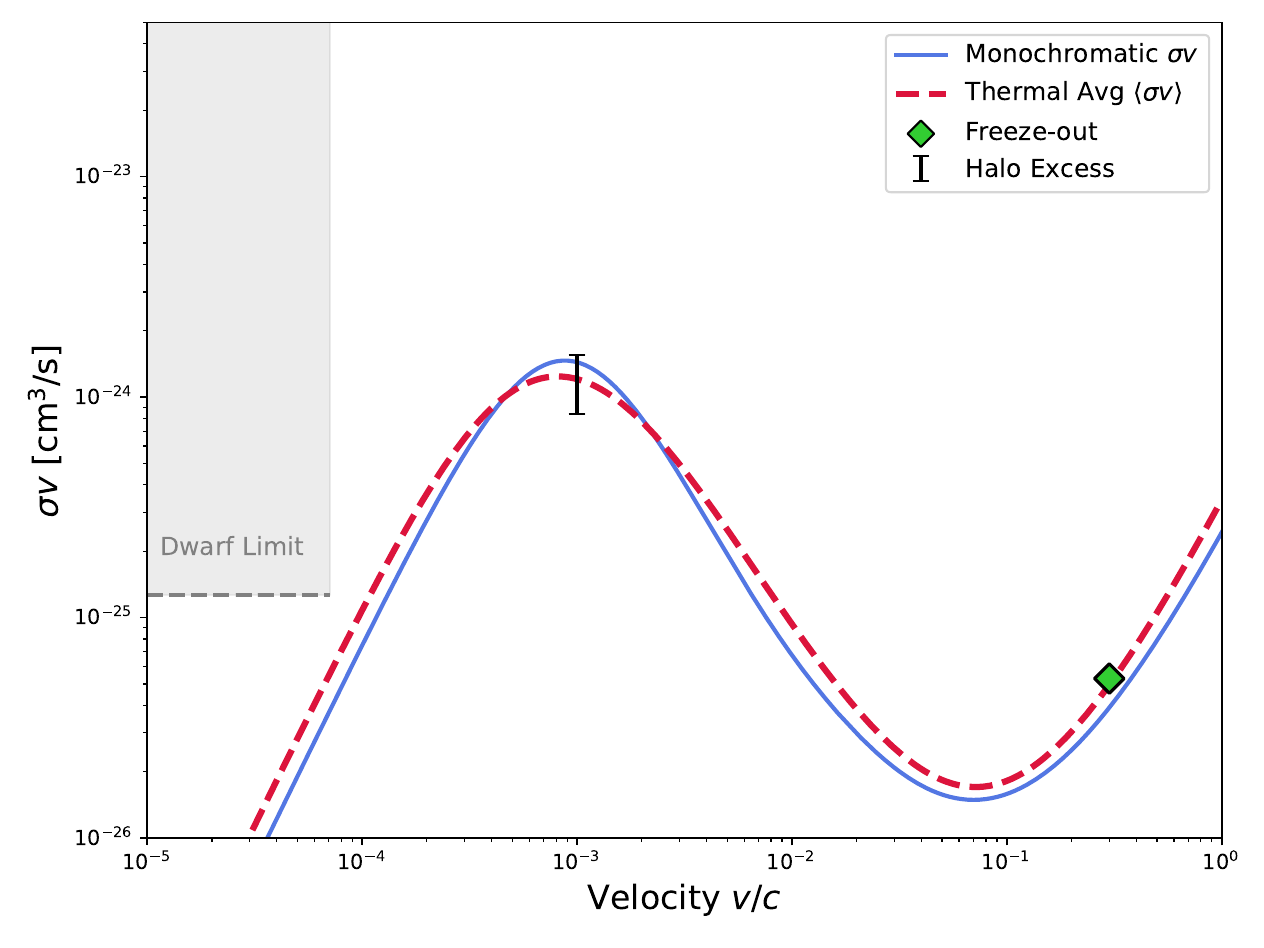}
\caption{Velocity dependence of the annihilation cross section with the illustrative parameter set given in the main text. We show the monochromatic (blue solid) and Maxwellian-weighted (red dashed) $\langle \sigma v \rangle$. Desired values for freeze-out (green marker) and Halo excess (black dot) cross sections are shown. The gray region shows the constraints from dwarf galaxy observations. }
\label{fig:xsec_velocity_dep}
\end{figure}

\begin{enumerate}
    \item \textbf{Perturbative Regime} ($\alpha_{\rm eff} \lesssim v < 1$): Here $S_l \approx 1$, and the cross section retains its standard $p$-wave scaling, $\langle \sigma v \rangle \propto v^2$.
    \item \textbf{Sommerfeld Regime} ($v_{\rm sat} \lesssim v \lesssim \alpha_{\rm eff}$): The $p$-wave SE scales as $v^{-3}$, yielding $\langle \sigma v \rangle \propto v^{-1}$. This regime provides the necessary boost to reproduce the observed $20$~GeV halo signal.
    \item \textbf{Saturation Regime} ($v \lesssim v_{\rm sat}$): Below the resonance-induced saturation velocity, $S_l$ becomes velocity-independent. Crucially, the cross section recovers its $\langle \sigma v \rangle \propto v^2$ suppression. This infrared ``shut-off'' ensures that the model satisfies stringent constraints from low-velocity environments, such as dwarf spheroidal galaxies~\cite{McDaniel:2023bju}.
\end{enumerate}

\section{Halo gamma-ray excess}

Based on the model established in the previous section, we calculate the differential photon spectrum resulting from the annihilation process $\chi \bar{\chi} \to \varphi \varphi$, followed by the subsequent decay $\varphi \to b \bar{b}$. The spectrum is evaluated using the \texttt{PPPC4DMID} package~\cite{Cirelli:2010xx}, accounting for the Lorentz boost of the mediator $\varphi$. This boost becomes particularly significant in the regime $m_\chi \gg m_\varphi$~\cite{Boddy:2016fds}, shifting the resulting photon distribution to higher energies in the laboratory frame. 

\begin{figure}[t]
\includegraphics[width=0.45\textwidth]{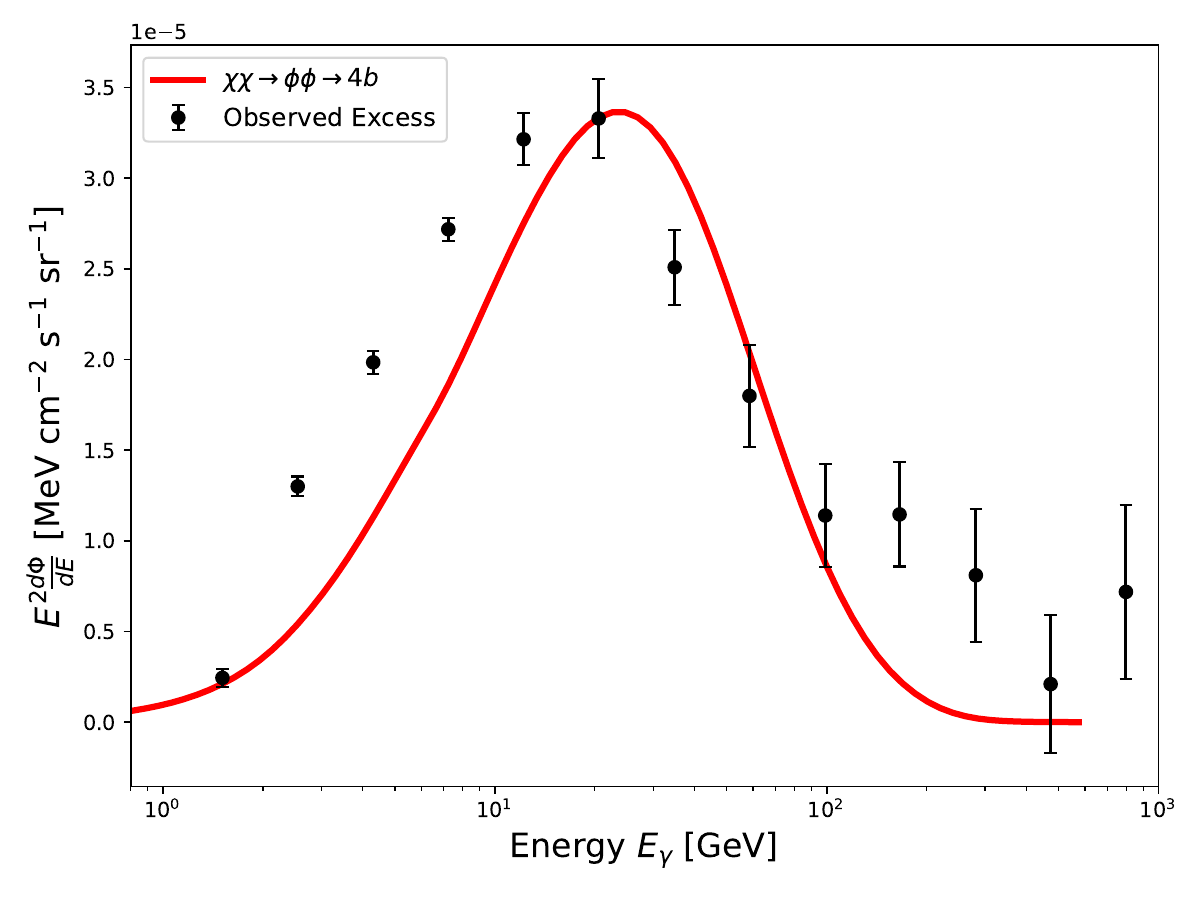}
\caption{$\langle \sigma v \rangle_{\rm Halo} \simeq 1.09 \times 10^{-24}$ cm$^3$/s for $\chi \bar{\chi} \to \varphi \varphi$. After the production of a pair of scalar $\varphi \varphi$ in on-shell, subsequent decays $\varphi \to b \bar{b}$ gives the gamma-ray spectrum including Lorentz boost effect \cite{Boddy:2016fds}.}
\label{fig:gamma_flux}
\end{figure}

We find that the $20~\text{GeV}$ gamma-ray excess is well-reproduced by a benchmark parameters:
\begin{align}
(m_\chi, m_\varphi, f_\varphi)= (575, 11.6, 873) ~{\rm GeV}
\end{align}
or the dark matter-mediator coupling is $y_\chi \equiv m_\chi/f_\varphi \simeq 1.518$ (corresponding to $\alpha_{\rm eff} \simeq 0.183$). For this parameter set, the velocity-averaged annihilation cross section in the Milky Way halo is $\langle \sigma v \rangle_{\rm Halo} \simeq 1.09 \times 10^{-24} \, \text{cm}^3/\text{s}$. The resulting signal spectrum, compared against the observed excess, is presented in Fig.~\ref{fig:gamma_flux}.

\section{Constraints}

\textit{Mediator-SM Coupling and Direct Detection}.---The dark sector communicates with the SM through Higgs-$\varphi$ mixing. For $m_\varphi \sim 10$~GeV, constraints from flavor-changing neutral currents (FCNCs) in $B$-meson decays typically require $\theta \lesssim 10^{-3}$--$10^{-4}$~\cite{Winkler:2018qyg}. However, for $m_\varphi \gtrsim 10$~GeV, these limits, along with those from rare quarkonia decays (e.g., $\Upsilon \to \gamma \varphi$), are significantly weakened or kinematically forbidden ($m_\varphi > m_{B,\Upsilon}$).

The most stringent upper bounds in this mass regime arise from spin-independent (SI) direct detection. The Higgs-portal interaction induces DM-nucleon scattering via $t$-channel mediator exchange. Based on the latest LUX-ZEPLIN (LZ)~\cite{LZ:2024zvo} and XENONnT~\cite{XENON:2025vwd} results, we obtain $\theta \lsim 5.6 \times 10^{-5}$ and $\theta \lsim 2.0 \times 10^{-4}$, respectively, for our benchmark parameters ($m_\chi = 575$~GeV, $m_\phi = 11.6$~GeV).

Lower bounds on the mixing angle are determined by early-universe cosmology. To ensure $\varphi$ decays prior to Big Bang Nucleosynthesis (BBN) ($t_\varphi < 1$~s) and that the DM maintained thermal equilibrium with the SM plasma before freeze-out, we require $\theta \gtrsim 10^{-7}$~\cite{Winkler:2018qyg}. Additionally, LHC searches for dimuon resonances ($pp \to \varphi \to \mu^+\mu^-$) currently allow $\sin\theta \lesssim 0.1$ in this mass range~\cite{Winkler:2018qyg, Haisch:2016hzu}. Integrating these constraints, we find a viable parameter window $10^{-7} \lesssim \theta \lesssim 10^{-5}$ for $m_\varphi \sim 11.6$~GeV, where the model simultaneously satisfies the halo excess and all laboratory bounds.

\textit{Antiproton Constraints}.---The production of multiple $b$-quarks in the final state yields a significant flux of cosmic-ray antiprotons. While recent AMS-02 data impose stringent limits on hadronic dark matter annihilation—particularly for direct $b\bar{b}$ final states~\cite{Berdugo:2022zzo}—these constraints are significantly alleviated by the kinematics of the cascade process $\chi\bar{\chi} \to \varphi\varphi \to 4b$. 
Because the mediator is light ($m_\varphi \sim 10$~GeV), the resulting antiproton spectrum is substantially softened compared to direct $2b$ annihilation~\cite{Wang:2025zth}. This spectral shift moves the peak of the antiproton signal toward lower energies where the astrophysical background is less certain and the detector sensitivity is also reduced. Furthermore, the non-negligible branching fraction into leptons ($\text{Br}(\varphi \to \tau^+\tau^-) \sim 16\%$) reduces the total hadronic yield. Consequently, the parameter space required to explain the 20~GeV halo excess remains fully compatible with current AMS-02 antiproton observations.

\section{Discussion and Conclusion}

In this Letter, we have demonstrated that the recently reported halo-like gamma-ray excess~\cite{Totani:2025fxx} can be naturally accommodated within a dark matter (DM) framework featuring $p$-wave annihilation and resonant Sommerfeld enhancement. The interplay between velocity-dependent $p$-wave suppression and resonant enhancement offers a robust mechanism for explaining emerging anomalies in the gamma-ray sky.  Our model successfully reconciles the large annihilation cross section required for the halo signal, $\langle \sigma v \rangle_{\text{Halo}} \sim 10^{-24}~\text{cm}^3/\text{s}$, with the standard thermal freeze-out relic abundance and the stringent null results from dwarf spheroidal galaxies.

It will be extremely interesting to find the light scalar $\varphi$ at future experiments. Given its small mixing angle $\theta$ and a mass in the $\mathcal {O} (10)$ GeV range, $\varphi$ typically manifests as a long-lived particle in collider environments \cite{Alimena:2019zri}, despite acting as a prompt decay source on galactic scales due to its negligible decay length. This light scalar portal can be effectively probed at the proposed SHiP facility \cite{Alekhin:2015byh}
and by dedicated displaced vertex detectors such as MATHUSLA \cite{Curtin:2018mvb} and/or the LHC timing detectors~\cite{Flowers:2019gvj}. Additionally, the forward physics facilities at the LHC, including FASER \cite{FASER:2019aik}, provide a unique opportunity to explore this specific mass regime. In the longer term, future $e^+e^-$ colliders like the FCC-ee \cite{FCC:2018evy} are expected to offer a clean experimental environment for the precision measurement of dark sector properties \cite{Blondel:2022qqo}.

\section*{Acknowledgments}
This work was supported by the National Research
Foundation of Korea (NRF) grant funded by the Korea government (MSIT) (RS-2024-00340153) and Yonsei internal grant for Mega-science (2023-22-048). 

\bigskip
\bibliographystyle{apsrev4-2}
\bibliography{ref}

\begin{thebibliography}{59}%
\makeatletter
\providecommand \@ifxundefined [1]{%
 \@ifx{#1\undefined}
}%
\providecommand \@ifnum [1]{%
 \ifnum #1\expandafter \@firstoftwo
 \else \expandafter \@secondoftwo
 \fi
}%
\providecommand \@ifx [1]{%
 \ifx #1\expandafter \@firstoftwo
 \else \expandafter \@secondoftwo
 \fi
}%
\providecommand \natexlab [1]{#1}%
\providecommand \enquote  [1]{``#1''}%
\providecommand \bibnamefont  [1]{#1}%
\providecommand \bibfnamefont [1]{#1}%
\providecommand \citenamefont [1]{#1}%
\providecommand \href@noop [0]{\@secondoftwo}%
\providecommand \href [0]{\begingroup \@sanitize@url \@href}%
\providecommand \@href[1]{\@@startlink{#1}\@@href}%
\providecommand \@@href[1]{\endgroup#1\@@endlink}%
\providecommand \@sanitize@url [0]{\catcode `\\12\catcode `\$12\catcode
  `\&12\catcode `\#12\catcode `\^12\catcode `\_12\catcode `\%12\relax}%
\providecommand \@@startlink[1]{}%
\providecommand \@@endlink[0]{}%
\providecommand \url  [0]{\begingroup\@sanitize@url \@url }%
\providecommand \@url [1]{\endgroup\@href {#1}{\urlprefix }}%
\providecommand \urlprefix  [0]{URL }%
\providecommand \Eprint [0]{\href }%
\providecommand \doibase [0]{https://doi.org/}%
\providecommand \selectlanguage [0]{\@gobble}%
\providecommand \bibinfo  [0]{\@secondoftwo}%
\providecommand \bibfield  [0]{\@secondoftwo}%
\providecommand \translation [1]{[#1]}%
\providecommand \BibitemOpen [0]{}%
\providecommand \bibitemStop [0]{}%
\providecommand \bibitemNoStop [0]{.\EOS\space}%
\providecommand \EOS [0]{\spacefactor3000\relax}%
\providecommand \BibitemShut  [1]{\csname bibitem#1\endcsname}%
\let\auto@bib@innerbib\@empty
\bibitem [{\citenamefont {Roszkowski}\ \emph {et~al.}(2018)\citenamefont
  {Roszkowski}, \citenamefont {Sessolo},\ and\ \citenamefont
  {Trojanowski}}]{Roszkowski:2017nbc}%
  \BibitemOpen
  \bibfield  {author} {\bibinfo {author} {\bibfnamefont {L.}~\bibnamefont
  {Roszkowski}}, \bibinfo {author} {\bibfnamefont {E.~M.}\ \bibnamefont
  {Sessolo}},\ and\ \bibinfo {author} {\bibfnamefont {S.}~\bibnamefont
  {Trojanowski}},\ }\href {https://doi.org/10.1088/1361-6633/aab913} {\bibfield
   {journal} {\bibinfo  {journal} {Rept. Prog. Phys.}\ }\textbf {\bibinfo
  {volume} {81}},\ \bibinfo {pages} {066201} (\bibinfo {year} {2018})},\
  \Eprint {https://arxiv.org/abs/1707.06277} {arXiv:1707.06277 [hep-ph]}
  \BibitemShut {NoStop}%
\bibitem [{\citenamefont {Abdollahi}\ \emph {et~al.}(2025)\citenamefont
  {Abdollahi} \emph {et~al.}}]{Fermi-LAT:2025gei}%
  \BibitemOpen
  \bibfield  {author} {\bibinfo {author} {\bibfnamefont {S.}~\bibnamefont
  {Abdollahi}} \emph {et~al.} (\bibinfo {collaboration} {Fermi-LAT, HAWC,
  H.E.S.S., MAGIC, VERITAS}),\ }\href@noop {} {\  (\bibinfo {year} {2025})},\
  \Eprint {https://arxiv.org/abs/2508.20229} {arXiv:2508.20229 [astro-ph.HE]}
  \BibitemShut {NoStop}%
\bibitem [{\citenamefont {Bulbul}\ \emph {et~al.}(2014)\citenamefont {Bulbul},
  \citenamefont {Markevitch}, \citenamefont {Foster}, \citenamefont {Smith},
  \citenamefont {Loewenstein},\ and\ \citenamefont {Randall}}]{Bulbul:2014sua}%
  \BibitemOpen
  \bibfield  {author} {\bibinfo {author} {\bibfnamefont {E.}~\bibnamefont
  {Bulbul}}, \bibinfo {author} {\bibfnamefont {M.}~\bibnamefont {Markevitch}},
  \bibinfo {author} {\bibfnamefont {A.}~\bibnamefont {Foster}}, \bibinfo
  {author} {\bibfnamefont {R.~K.}\ \bibnamefont {Smith}}, \bibinfo {author}
  {\bibfnamefont {M.}~\bibnamefont {Loewenstein}},\ and\ \bibinfo {author}
  {\bibfnamefont {S.~W.}\ \bibnamefont {Randall}},\ }\href
  {https://doi.org/10.1088/0004-637X/789/1/13} {\bibfield  {journal} {\bibinfo
  {journal} {Astrophys. J.}\ }\textbf {\bibinfo {volume} {789}},\ \bibinfo
  {pages} {13} (\bibinfo {year} {2014})},\ \Eprint
  {https://arxiv.org/abs/1402.2301} {arXiv:1402.2301 [astro-ph.CO]}
  \BibitemShut {NoStop}%
\bibitem [{\citenamefont {Boyarsky}\ \emph {et~al.}(2014)\citenamefont
  {Boyarsky}, \citenamefont {Ruchayskiy}, \citenamefont {Iakubovskyi},\ and\
  \citenamefont {Franse}}]{Boyarsky:2014jta}%
  \BibitemOpen
  \bibfield  {author} {\bibinfo {author} {\bibfnamefont {A.}~\bibnamefont
  {Boyarsky}}, \bibinfo {author} {\bibfnamefont {O.}~\bibnamefont
  {Ruchayskiy}}, \bibinfo {author} {\bibfnamefont {D.}~\bibnamefont
  {Iakubovskyi}},\ and\ \bibinfo {author} {\bibfnamefont {J.}~\bibnamefont
  {Franse}},\ }\href {https://doi.org/10.1103/PhysRevLett.113.251301}
  {\bibfield  {journal} {\bibinfo  {journal} {Phys. Rev. Lett.}\ }\textbf
  {\bibinfo {volume} {113}},\ \bibinfo {pages} {251301} (\bibinfo {year}
  {2014})},\ \Eprint {https://arxiv.org/abs/1402.4119} {arXiv:1402.4119
  [astro-ph.CO]} \BibitemShut {NoStop}%
\bibitem [{\citenamefont {Teegarden}\ \emph {et~al.}(2005)\citenamefont
  {Teegarden} \emph {et~al.}}]{Teegarden:2004ct}%
  \BibitemOpen
  \bibfield  {author} {\bibinfo {author} {\bibfnamefont {B.~J.}\ \bibnamefont
  {Teegarden}} \emph {et~al.},\ }\href {https://doi.org/10.1086/426859}
  {\bibfield  {journal} {\bibinfo  {journal} {Astrophys. J.}\ }\textbf
  {\bibinfo {volume} {621}},\ \bibinfo {pages} {296} (\bibinfo {year}
  {2005})},\ \Eprint {https://arxiv.org/abs/astro-ph/0410354}
  {arXiv:astro-ph/0410354} \BibitemShut {NoStop}%
\bibitem [{\citenamefont {Knodlseder}\ \emph {et~al.}(2005)\citenamefont
  {Knodlseder} \emph {et~al.}}]{Knodlseder:2005yq}%
  \BibitemOpen
  \bibfield  {author} {\bibinfo {author} {\bibfnamefont {J.}~\bibnamefont
  {Knodlseder}} \emph {et~al.},\ }\href
  {https://doi.org/10.1051/0004-6361:20042063} {\bibfield  {journal} {\bibinfo
  {journal} {Astron. Astrophys.}\ }\textbf {\bibinfo {volume} {441}},\ \bibinfo
  {pages} {513} (\bibinfo {year} {2005})},\ \Eprint
  {https://arxiv.org/abs/astro-ph/0506026} {arXiv:astro-ph/0506026}
  \BibitemShut {NoStop}%
\bibitem [{\citenamefont {Weniger}(2012)}]{Weniger:2012tx}%
  \BibitemOpen
  \bibfield  {author} {\bibinfo {author} {\bibfnamefont {C.}~\bibnamefont
  {Weniger}},\ }\href {https://doi.org/10.1088/1475-7516/2012/08/007}
  {\bibfield  {journal} {\bibinfo  {journal} {JCAP}\ }\textbf {\bibinfo
  {volume} {08}},\ \bibinfo {pages} {007}},\ \Eprint
  {https://arxiv.org/abs/1204.2797} {arXiv:1204.2797 [hep-ph]} \BibitemShut
  {NoStop}%
\bibitem [{\citenamefont {Bringmann}\ \emph {et~al.}(2012)\citenamefont
  {Bringmann}, \citenamefont {Huang}, \citenamefont {Ibarra}, \citenamefont
  {Vogl},\ and\ \citenamefont {Weniger}}]{Bringmann:2012vr}%
  \BibitemOpen
  \bibfield  {author} {\bibinfo {author} {\bibfnamefont {T.}~\bibnamefont
  {Bringmann}}, \bibinfo {author} {\bibfnamefont {X.}~\bibnamefont {Huang}},
  \bibinfo {author} {\bibfnamefont {A.}~\bibnamefont {Ibarra}}, \bibinfo
  {author} {\bibfnamefont {S.}~\bibnamefont {Vogl}},\ and\ \bibinfo {author}
  {\bibfnamefont {C.}~\bibnamefont {Weniger}},\ }\href
  {https://doi.org/10.1088/1475-7516/2012/07/054} {\bibfield  {journal}
  {\bibinfo  {journal} {JCAP}\ }\textbf {\bibinfo {volume} {07}},\ \bibinfo
  {pages} {054}},\ \Eprint {https://arxiv.org/abs/1203.1312} {arXiv:1203.1312
  [hep-ph]} \BibitemShut {NoStop}%
\bibitem [{\citenamefont {Huh}\ \emph {et~al.}(2008)\citenamefont {Huh},
  \citenamefont {Kim}, \citenamefont {Park},\ and\ \citenamefont
  {Park}}]{Huh:2007zw}%
  \BibitemOpen
  \bibfield  {author} {\bibinfo {author} {\bibfnamefont {J.-H.}\ \bibnamefont
  {Huh}}, \bibinfo {author} {\bibfnamefont {J.~E.}\ \bibnamefont {Kim}},
  \bibinfo {author} {\bibfnamefont {J.-C.}\ \bibnamefont {Park}},\ and\
  \bibinfo {author} {\bibfnamefont {S.~C.}\ \bibnamefont {Park}},\ }\href
  {https://doi.org/10.1103/PhysRevD.77.123503} {\bibfield  {journal} {\bibinfo
  {journal} {Phys. Rev. D}\ }\textbf {\bibinfo {volume} {77}},\ \bibinfo
  {pages} {123503} (\bibinfo {year} {2008})},\ \Eprint
  {https://arxiv.org/abs/0711.3528} {arXiv:0711.3528 [astro-ph]} \BibitemShut
  {NoStop}%
\bibitem [{\citenamefont {Park}\ and\ \citenamefont
  {Park}(2013)}]{Park:2012xq}%
  \BibitemOpen
  \bibfield  {author} {\bibinfo {author} {\bibfnamefont {J.-C.}\ \bibnamefont
  {Park}}\ and\ \bibinfo {author} {\bibfnamefont {S.~C.}\ \bibnamefont
  {Park}},\ }\href {https://doi.org/10.1016/j.physletb.2012.12.035} {\bibfield
  {journal} {\bibinfo  {journal} {Phys. Lett. B}\ }\textbf {\bibinfo {volume}
  {718}},\ \bibinfo {pages} {1401} (\bibinfo {year} {2013})},\ \Eprint
  {https://arxiv.org/abs/1207.4981} {arXiv:1207.4981 [hep-ph]} \BibitemShut
  {NoStop}%
\bibitem [{\citenamefont {Park}\ \emph {et~al.}(2014)\citenamefont {Park},
  \citenamefont {Park},\ and\ \citenamefont {Kong}}]{Park:2014olo}%
  \BibitemOpen
  \bibfield  {author} {\bibinfo {author} {\bibfnamefont {J.-C.}\ \bibnamefont
  {Park}}, \bibinfo {author} {\bibfnamefont {S.~C.}\ \bibnamefont {Park}},\
  and\ \bibinfo {author} {\bibfnamefont {K.}~\bibnamefont {Kong}},\ }\href
  {https://doi.org/10.1016/j.physletb.2014.04.037} {\bibfield  {journal}
  {\bibinfo  {journal} {Phys. Lett. B}\ }\textbf {\bibinfo {volume} {733}},\
  \bibinfo {pages} {217} (\bibinfo {year} {2014})},\ \Eprint
  {https://arxiv.org/abs/1403.1536} {arXiv:1403.1536 [hep-ph]} \BibitemShut
  {NoStop}%
\bibitem [{\citenamefont {Lee}\ \emph {et~al.}(2014)\citenamefont {Lee},
  \citenamefont {Park},\ and\ \citenamefont {Park}}]{Lee:2014xua}%
  \BibitemOpen
  \bibfield  {author} {\bibinfo {author} {\bibfnamefont {H.~M.}\ \bibnamefont
  {Lee}}, \bibinfo {author} {\bibfnamefont {S.~C.}\ \bibnamefont {Park}},\ and\
  \bibinfo {author} {\bibfnamefont {W.-I.}\ \bibnamefont {Park}},\ }\href
  {https://doi.org/10.1140/epjc/s10052-014-3062-5} {\bibfield  {journal}
  {\bibinfo  {journal} {Eur. Phys. J. C}\ }\textbf {\bibinfo {volume} {74}},\
  \bibinfo {pages} {3062} (\bibinfo {year} {2014})},\ \Eprint
  {https://arxiv.org/abs/1403.0865} {arXiv:1403.0865 [astro-ph.CO]}
  \BibitemShut {NoStop}%
\bibitem [{\citenamefont {Park}\ \emph {et~al.}(2016)\citenamefont {Park},
  \citenamefont {Kim},\ and\ \citenamefont {Park}}]{Park:2015gdo}%
  \BibitemOpen
  \bibfield  {author} {\bibinfo {author} {\bibfnamefont {J.-C.}\ \bibnamefont
  {Park}}, \bibinfo {author} {\bibfnamefont {J.}~\bibnamefont {Kim}},\ and\
  \bibinfo {author} {\bibfnamefont {S.~C.}\ \bibnamefont {Park}},\ }\href
  {https://doi.org/10.1016/j.physletb.2015.11.035} {\bibfield  {journal}
  {\bibinfo  {journal} {Phys. Lett. B}\ }\textbf {\bibinfo {volume} {752}},\
  \bibinfo {pages} {59} (\bibinfo {year} {2016})},\ \Eprint
  {https://arxiv.org/abs/1505.04620} {arXiv:1505.04620 [hep-ph]} \BibitemShut
  {NoStop}%
\bibitem [{\citenamefont {Park}\ and\ \citenamefont
  {Park}(2016)}]{Park:2015ysf}%
  \BibitemOpen
  \bibfield  {author} {\bibinfo {author} {\bibfnamefont {J.-C.}\ \bibnamefont
  {Park}}\ and\ \bibinfo {author} {\bibfnamefont {S.~C.}\ \bibnamefont
  {Park}},\ }\href {https://doi.org/10.1016/j.dark.2016.08.002} {\bibfield
  {journal} {\bibinfo  {journal} {Phys. Dark Univ.}\ }\textbf {\bibinfo
  {volume} {14}},\ \bibinfo {pages} {4} (\bibinfo {year} {2016})},\ \Eprint
  {https://arxiv.org/abs/1512.08117} {arXiv:1512.08117 [hep-ph]} \BibitemShut
  {NoStop}%
\bibitem [{\citenamefont {Ibarra}\ \emph {et~al.}(2013)\citenamefont {Ibarra},
  \citenamefont {Tran},\ and\ \citenamefont {Weniger}}]{Ibarra:2013cra}%
  \BibitemOpen
  \bibfield  {author} {\bibinfo {author} {\bibfnamefont {A.}~\bibnamefont
  {Ibarra}}, \bibinfo {author} {\bibfnamefont {D.}~\bibnamefont {Tran}},\ and\
  \bibinfo {author} {\bibfnamefont {C.}~\bibnamefont {Weniger}},\ }\href
  {https://doi.org/10.1142/S0217751X13300408} {\bibfield  {journal} {\bibinfo
  {journal} {Int. J. Mod. Phys. A}\ }\textbf {\bibinfo {volume} {28}},\
  \bibinfo {pages} {1330040} (\bibinfo {year} {2013})},\ \Eprint
  {https://arxiv.org/abs/1307.6434} {arXiv:1307.6434 [hep-ph]} \BibitemShut
  {NoStop}%
\bibitem [{\citenamefont {Cline}\ and\ \citenamefont
  {Frey}(2014)}]{Cline:2014vsa}%
  \BibitemOpen
  \bibfield  {author} {\bibinfo {author} {\bibfnamefont {J.~M.}\ \bibnamefont
  {Cline}}\ and\ \bibinfo {author} {\bibfnamefont {A.~R.}\ \bibnamefont
  {Frey}},\ }\href {https://doi.org/10.1103/PhysRevD.90.123537} {\bibfield
  {journal} {\bibinfo  {journal} {Phys. Rev. D}\ }\textbf {\bibinfo {volume}
  {90}},\ \bibinfo {pages} {123537} (\bibinfo {year} {2014})},\ \Eprint
  {https://arxiv.org/abs/1410.7766} {arXiv:1410.7766 [astro-ph.CO]}
  \BibitemShut {NoStop}%
\bibitem [{\citenamefont {Totani}(2025)}]{Totani:2025fxx}%
  \BibitemOpen
  \bibfield  {author} {\bibinfo {author} {\bibfnamefont {T.}~\bibnamefont
  {Totani}},\ }\href {https://doi.org/10.1088/1475-7516/2025/11/080} {\bibfield
   {journal} {\bibinfo  {journal} {JCAP}\ }\textbf {\bibinfo {volume} {11}},\
  \bibinfo {pages} {080}},\ \Eprint {https://arxiv.org/abs/2507.07209}
  {arXiv:2507.07209 [astro-ph.HE]} \BibitemShut {NoStop}%
\bibitem [{\citenamefont {Murayama}(2025)}]{Murayama:2025ihg}%
  \BibitemOpen
  \bibfield  {author} {\bibinfo {author} {\bibfnamefont {H.}~\bibnamefont
  {Murayama}},\ }\href@noop {} {\  (\bibinfo {year} {2025})},\ \Eprint
  {https://arxiv.org/abs/2512.01404} {arXiv:2512.01404 [hep-ph]} \BibitemShut
  {NoStop}%
\bibitem [{\citenamefont {Choquette}\ \emph {et~al.}(2016)\citenamefont
  {Choquette}, \citenamefont {Cline},\ and\ \citenamefont
  {Cornell}}]{Choquette:2016xsw}%
  \BibitemOpen
  \bibfield  {author} {\bibinfo {author} {\bibfnamefont {J.}~\bibnamefont
  {Choquette}}, \bibinfo {author} {\bibfnamefont {J.~M.}\ \bibnamefont
  {Cline}},\ and\ \bibinfo {author} {\bibfnamefont {J.~M.}\ \bibnamefont
  {Cornell}},\ }\href {https://doi.org/10.1103/PhysRevD.94.015018} {\bibfield
  {journal} {\bibinfo  {journal} {Phys. Rev. D}\ }\textbf {\bibinfo {volume}
  {94}},\ \bibinfo {pages} {015018} (\bibinfo {year} {2016})},\ \Eprint
  {https://arxiv.org/abs/1604.01039} {arXiv:1604.01039 [hep-ph]} \BibitemShut
  {NoStop}%
\bibitem [{\citenamefont {Walker}\ \emph {et~al.}(2007)\citenamefont {Walker},
  \citenamefont {Mateo}, \citenamefont {Olszewski}, \citenamefont {Gnedin},
  \citenamefont {Wang}, \citenamefont {Sen},\ and\ \citenamefont
  {Woodroofe}}]{Walker:2007ju}%
  \BibitemOpen
  \bibfield  {author} {\bibinfo {author} {\bibfnamefont {M.~G.}\ \bibnamefont
  {Walker}}, \bibinfo {author} {\bibfnamefont {M.}~\bibnamefont {Mateo}},
  \bibinfo {author} {\bibfnamefont {E.~W.}\ \bibnamefont {Olszewski}}, \bibinfo
  {author} {\bibfnamefont {O.~Y.}\ \bibnamefont {Gnedin}}, \bibinfo {author}
  {\bibfnamefont {X.}~\bibnamefont {Wang}}, \bibinfo {author} {\bibfnamefont
  {B.}~\bibnamefont {Sen}},\ and\ \bibinfo {author} {\bibfnamefont
  {M.}~\bibnamefont {Woodroofe}},\ }\href {https://doi.org/10.1086/521998}
  {\bibfield  {journal} {\bibinfo  {journal} {Astrophys. J. Lett.}\ }\textbf
  {\bibinfo {volume} {667}},\ \bibinfo {pages} {L53} (\bibinfo {year}
  {2007})},\ \Eprint {https://arxiv.org/abs/0708.0010} {arXiv:0708.0010
  [astro-ph]} \BibitemShut {NoStop}%
\bibitem [{\citenamefont {Dehnen}\ \emph {et~al.}(2006)\citenamefont {Dehnen},
  \citenamefont {McLaughlin},\ and\ \citenamefont {Sachania}}]{Dehnen:2006cm}%
  \BibitemOpen
  \bibfield  {author} {\bibinfo {author} {\bibfnamefont {W.}~\bibnamefont
  {Dehnen}}, \bibinfo {author} {\bibfnamefont {D.}~\bibnamefont {McLaughlin}},\
  and\ \bibinfo {author} {\bibfnamefont {J.}~\bibnamefont {Sachania}},\ }\href
  {https://doi.org/10.1111/j.1365-2966.2006.10404.x} {\bibfield  {journal}
  {\bibinfo  {journal} {Mon. Not. Roy. Astron. Soc.}\ }\textbf {\bibinfo
  {volume} {369}},\ \bibinfo {pages} {1688} (\bibinfo {year} {2006})},\ \Eprint
  {https://arxiv.org/abs/astro-ph/0603825} {arXiv:astro-ph/0603825}
  \BibitemShut {NoStop}%
\bibitem [{\citenamefont {Sommerfeld}(1931)}]{Sommerfeld:1931qaf}%
  \BibitemOpen
  \bibfield  {author} {\bibinfo {author} {\bibfnamefont {A.}~\bibnamefont
  {Sommerfeld}},\ }\href {https://doi.org/10.1002/andp.19314030302} {\bibfield
  {journal} {\bibinfo  {journal} {Annalen Phys.}\ }\textbf {\bibinfo {volume}
  {403}},\ \bibinfo {pages} {257} (\bibinfo {year} {1931})}\BibitemShut
  {NoStop}%
\bibitem [{\citenamefont {Baer}\ \emph {et~al.}(1999)\citenamefont {Baer},
  \citenamefont {Cheung},\ and\ \citenamefont {Gunion}}]{Baer:1998pg}%
  \BibitemOpen
  \bibfield  {author} {\bibinfo {author} {\bibfnamefont {H.}~\bibnamefont
  {Baer}}, \bibinfo {author} {\bibfnamefont {K.-m.}\ \bibnamefont {Cheung}},\
  and\ \bibinfo {author} {\bibfnamefont {J.~F.}\ \bibnamefont {Gunion}},\
  }\href {https://doi.org/10.1103/PhysRevD.59.075002} {\bibfield  {journal}
  {\bibinfo  {journal} {Phys. Rev. D}\ }\textbf {\bibinfo {volume} {59}},\
  \bibinfo {pages} {075002} (\bibinfo {year} {1999})},\ \Eprint
  {https://arxiv.org/abs/hep-ph/9806361} {arXiv:hep-ph/9806361} \BibitemShut
  {NoStop}%
\bibitem [{\citenamefont {Hisano}\ \emph {et~al.}(2004)\citenamefont {Hisano},
  \citenamefont {Matsumoto},\ and\ \citenamefont {Nojiri}}]{Hisano:2003ec}%
  \BibitemOpen
  \bibfield  {author} {\bibinfo {author} {\bibfnamefont {J.}~\bibnamefont
  {Hisano}}, \bibinfo {author} {\bibfnamefont {S.}~\bibnamefont {Matsumoto}},\
  and\ \bibinfo {author} {\bibfnamefont {M.~M.}\ \bibnamefont {Nojiri}},\
  }\href {https://doi.org/10.1103/PhysRevLett.92.031303} {\bibfield  {journal}
  {\bibinfo  {journal} {Phys. Rev. Lett.}\ }\textbf {\bibinfo {volume} {92}},\
  \bibinfo {pages} {031303} (\bibinfo {year} {2004})},\ \Eprint
  {https://arxiv.org/abs/hep-ph/0307216} {arXiv:hep-ph/0307216} \BibitemShut
  {NoStop}%
\bibitem [{\citenamefont {Hisano}\ \emph {et~al.}(2005)\citenamefont {Hisano},
  \citenamefont {Matsumoto}, \citenamefont {Nojiri},\ and\ \citenamefont
  {Saito}}]{Hisano:2004ds}%
  \BibitemOpen
  \bibfield  {author} {\bibinfo {author} {\bibfnamefont {J.}~\bibnamefont
  {Hisano}}, \bibinfo {author} {\bibfnamefont {S.}~\bibnamefont {Matsumoto}},
  \bibinfo {author} {\bibfnamefont {M.~M.}\ \bibnamefont {Nojiri}},\ and\
  \bibinfo {author} {\bibfnamefont {O.}~\bibnamefont {Saito}},\ }\href
  {https://doi.org/10.1103/PhysRevD.71.063528} {\bibfield  {journal} {\bibinfo
  {journal} {Phys. Rev. D}\ }\textbf {\bibinfo {volume} {71}},\ \bibinfo
  {pages} {063528} (\bibinfo {year} {2005})},\ \Eprint
  {https://arxiv.org/abs/hep-ph/0412403} {arXiv:hep-ph/0412403} \BibitemShut
  {NoStop}%
\bibitem [{\citenamefont {Hisano}\ \emph {et~al.}(2006)\citenamefont {Hisano},
  \citenamefont {Matsumoto}, \citenamefont {Saito},\ and\ \citenamefont
  {Senami}}]{Hisano:2005ec}%
  \BibitemOpen
  \bibfield  {author} {\bibinfo {author} {\bibfnamefont {J.}~\bibnamefont
  {Hisano}}, \bibinfo {author} {\bibfnamefont {S.}~\bibnamefont {Matsumoto}},
  \bibinfo {author} {\bibfnamefont {O.}~\bibnamefont {Saito}},\ and\ \bibinfo
  {author} {\bibfnamefont {M.}~\bibnamefont {Senami}},\ }\href
  {https://doi.org/10.1103/PhysRevD.73.055004} {\bibfield  {journal} {\bibinfo
  {journal} {Phys. Rev. D}\ }\textbf {\bibinfo {volume} {73}},\ \bibinfo
  {pages} {055004} (\bibinfo {year} {2006})},\ \Eprint
  {https://arxiv.org/abs/hep-ph/0511118} {arXiv:hep-ph/0511118} \BibitemShut
  {NoStop}%
\bibitem [{\citenamefont {Cirelli}\ \emph {et~al.}(2007)\citenamefont
  {Cirelli}, \citenamefont {Strumia},\ and\ \citenamefont
  {Tamburini}}]{Cirelli:2007xd}%
  \BibitemOpen
  \bibfield  {author} {\bibinfo {author} {\bibfnamefont {M.}~\bibnamefont
  {Cirelli}}, \bibinfo {author} {\bibfnamefont {A.}~\bibnamefont {Strumia}},\
  and\ \bibinfo {author} {\bibfnamefont {M.}~\bibnamefont {Tamburini}},\ }\href
  {https://doi.org/10.1016/j.nuclphysb.2007.07.023} {\bibfield  {journal}
  {\bibinfo  {journal} {Nucl. Phys. B}\ }\textbf {\bibinfo {volume} {787}},\
  \bibinfo {pages} {152} (\bibinfo {year} {2007})},\ \Eprint
  {https://arxiv.org/abs/0706.4071} {arXiv:0706.4071 [hep-ph]} \BibitemShut
  {NoStop}%
\bibitem [{\citenamefont {Cirelli}\ \emph {et~al.}(2009)\citenamefont
  {Cirelli}, \citenamefont {Kadastik}, \citenamefont {Raidal},\ and\
  \citenamefont {Strumia}}]{Cirelli:2008pk}%
  \BibitemOpen
  \bibfield  {author} {\bibinfo {author} {\bibfnamefont {M.}~\bibnamefont
  {Cirelli}}, \bibinfo {author} {\bibfnamefont {M.}~\bibnamefont {Kadastik}},
  \bibinfo {author} {\bibfnamefont {M.}~\bibnamefont {Raidal}},\ and\ \bibinfo
  {author} {\bibfnamefont {A.}~\bibnamefont {Strumia}},\ }\href
  {https://doi.org/10.1016/j.nuclphysb.2008.11.031} {\bibfield  {journal}
  {\bibinfo  {journal} {Nucl. Phys. B}\ }\textbf {\bibinfo {volume} {813}},\
  \bibinfo {pages} {1} (\bibinfo {year} {2009})},\ \bibinfo {note} {[Addendum:
  Nucl.Phys.B 873, 530--533 (2013)]},\ \Eprint
  {https://arxiv.org/abs/0809.2409} {arXiv:0809.2409 [hep-ph]} \BibitemShut
  {NoStop}%
\bibitem [{\citenamefont {Arkani-Hamed}\ \emph {et~al.}(2009)\citenamefont
  {Arkani-Hamed}, \citenamefont {Finkbeiner}, \citenamefont {Slatyer},\ and\
  \citenamefont {Weiner}}]{Arkani-Hamed:2008hhe}%
  \BibitemOpen
  \bibfield  {author} {\bibinfo {author} {\bibfnamefont {N.}~\bibnamefont
  {Arkani-Hamed}}, \bibinfo {author} {\bibfnamefont {D.~P.}\ \bibnamefont
  {Finkbeiner}}, \bibinfo {author} {\bibfnamefont {T.~R.}\ \bibnamefont
  {Slatyer}},\ and\ \bibinfo {author} {\bibfnamefont {N.}~\bibnamefont
  {Weiner}},\ }\href {https://doi.org/10.1103/PhysRevD.79.015014} {\bibfield
  {journal} {\bibinfo  {journal} {Phys. Rev. D}\ }\textbf {\bibinfo {volume}
  {79}},\ \bibinfo {pages} {015014} (\bibinfo {year} {2009})},\ \Eprint
  {https://arxiv.org/abs/0810.0713} {arXiv:0810.0713 [hep-ph]} \BibitemShut
  {NoStop}%
\bibitem [{\citenamefont {Pospelov}\ and\ \citenamefont
  {Ritz}(2009)}]{Pospelov:2008jd}%
  \BibitemOpen
  \bibfield  {author} {\bibinfo {author} {\bibfnamefont {M.}~\bibnamefont
  {Pospelov}}\ and\ \bibinfo {author} {\bibfnamefont {A.}~\bibnamefont
  {Ritz}},\ }\href {https://doi.org/10.1016/j.physletb.2008.12.012} {\bibfield
  {journal} {\bibinfo  {journal} {Phys. Lett. B}\ }\textbf {\bibinfo {volume}
  {671}},\ \bibinfo {pages} {391} (\bibinfo {year} {2009})},\ \Eprint
  {https://arxiv.org/abs/0810.1502} {arXiv:0810.1502 [hep-ph]} \BibitemShut
  {NoStop}%
\bibitem [{\citenamefont {Fox}\ and\ \citenamefont
  {Poppitz}(2009)}]{Fox:2008kb}%
  \BibitemOpen
  \bibfield  {author} {\bibinfo {author} {\bibfnamefont {P.~J.}\ \bibnamefont
  {Fox}}\ and\ \bibinfo {author} {\bibfnamefont {E.}~\bibnamefont {Poppitz}},\
  }\href {https://doi.org/10.1103/PhysRevD.79.083528} {\bibfield  {journal}
  {\bibinfo  {journal} {Phys. Rev. D}\ }\textbf {\bibinfo {volume} {79}},\
  \bibinfo {pages} {083528} (\bibinfo {year} {2009})},\ \Eprint
  {https://arxiv.org/abs/0811.0399} {arXiv:0811.0399 [hep-ph]} \BibitemShut
  {NoStop}%
\bibitem [{\citenamefont {Iengo}(2009)}]{Iengo:2009ni}%
  \BibitemOpen
  \bibfield  {author} {\bibinfo {author} {\bibfnamefont {R.}~\bibnamefont
  {Iengo}},\ }\href {https://doi.org/10.1088/1126-6708/2009/05/024} {\bibfield
  {journal} {\bibinfo  {journal} {JHEP}\ }\textbf {\bibinfo {volume} {05}},\
  \bibinfo {pages} {024}},\ \Eprint {https://arxiv.org/abs/0902.0688}
  {arXiv:0902.0688 [hep-ph]} \BibitemShut {NoStop}%
\bibitem [{\citenamefont {Feng}\ \emph {et~al.}(2010)\citenamefont {Feng},
  \citenamefont {Kaplinghat},\ and\ \citenamefont {Yu}}]{Feng:2010zp}%
  \BibitemOpen
  \bibfield  {author} {\bibinfo {author} {\bibfnamefont {J.~L.}\ \bibnamefont
  {Feng}}, \bibinfo {author} {\bibfnamefont {M.}~\bibnamefont {Kaplinghat}},\
  and\ \bibinfo {author} {\bibfnamefont {H.-B.}\ \bibnamefont {Yu}},\ }\href
  {https://doi.org/10.1103/PhysRevD.82.083525} {\bibfield  {journal} {\bibinfo
  {journal} {Phys. Rev. D}\ }\textbf {\bibinfo {volume} {82}},\ \bibinfo
  {pages} {083525} (\bibinfo {year} {2010})},\ \Eprint
  {https://arxiv.org/abs/1005.4678} {arXiv:1005.4678 [hep-ph]} \BibitemShut
  {NoStop}%
\bibitem [{\citenamefont {Coleman}\ and\ \citenamefont
  {Weinberg}(1973)}]{Coleman:1973jx}%
  \BibitemOpen
  \bibfield  {author} {\bibinfo {author} {\bibfnamefont {S.~R.}\ \bibnamefont
  {Coleman}}\ and\ \bibinfo {author} {\bibfnamefont {E.~J.}\ \bibnamefont
  {Weinberg}},\ }\href {https://doi.org/10.1103/PhysRevD.7.1888} {\bibfield
  {journal} {\bibinfo  {journal} {Phys. Rev. D}\ }\textbf {\bibinfo {volume}
  {7}},\ \bibinfo {pages} {1888} (\bibinfo {year} {1973})}\BibitemShut
  {NoStop}%
\bibitem [{\citenamefont {Gildener}\ and\ \citenamefont
  {Weinberg}(1976)}]{Gildener:1976ih}%
  \BibitemOpen
  \bibfield  {author} {\bibinfo {author} {\bibfnamefont {E.}~\bibnamefont
  {Gildener}}\ and\ \bibinfo {author} {\bibfnamefont {S.}~\bibnamefont
  {Weinberg}},\ }\href {https://doi.org/10.1103/PhysRevD.13.3333} {\bibfield
  {journal} {\bibinfo  {journal} {Phys. Rev. D}\ }\textbf {\bibinfo {volume}
  {13}},\ \bibinfo {pages} {3333} (\bibinfo {year} {1976})}\BibitemShut
  {NoStop}%
\bibitem [{\citenamefont {Brivio}\ and\ \citenamefont
  {Trott}(2019)}]{Brivio:2017vri}%
  \BibitemOpen
  \bibfield  {author} {\bibinfo {author} {\bibfnamefont {I.}~\bibnamefont
  {Brivio}}\ and\ \bibinfo {author} {\bibfnamefont {M.}~\bibnamefont {Trott}},\
  }\href {https://doi.org/10.1016/j.physrep.2018.11.002} {\bibfield  {journal}
  {\bibinfo  {journal} {Phys. Rept.}\ }\textbf {\bibinfo {volume} {793}},\
  \bibinfo {pages} {1} (\bibinfo {year} {2019})},\ \Eprint
  {https://arxiv.org/abs/1706.08945} {arXiv:1706.08945 [hep-ph]} \BibitemShut
  {NoStop}%
\bibitem [{\citenamefont {Bellazzini}\ \emph {et~al.}(2013)\citenamefont
  {Bellazzini}, \citenamefont {Csaki}, \citenamefont {Hubisz}, \citenamefont
  {Serra},\ and\ \citenamefont {Terning}}]{Bellazzini:2012vz}%
  \BibitemOpen
  \bibfield  {author} {\bibinfo {author} {\bibfnamefont {B.}~\bibnamefont
  {Bellazzini}}, \bibinfo {author} {\bibfnamefont {C.}~\bibnamefont {Csaki}},
  \bibinfo {author} {\bibfnamefont {J.}~\bibnamefont {Hubisz}}, \bibinfo
  {author} {\bibfnamefont {J.}~\bibnamefont {Serra}},\ and\ \bibinfo {author}
  {\bibfnamefont {J.}~\bibnamefont {Terning}},\ }\href
  {https://doi.org/10.1140/epjc/s10052-013-2333-x} {\bibfield  {journal}
  {\bibinfo  {journal} {Eur. Phys. J. C}\ }\textbf {\bibinfo {volume} {73}},\
  \bibinfo {pages} {2333} (\bibinfo {year} {2013})},\ \Eprint
  {https://arxiv.org/abs/1209.3299} {arXiv:1209.3299 [hep-ph]} \BibitemShut
  {NoStop}%
\bibitem [{\citenamefont {Bellazzini}\ \emph {et~al.}(2014)\citenamefont
  {Bellazzini}, \citenamefont {Csaki}, \citenamefont {Hubisz}, \citenamefont
  {Serra},\ and\ \citenamefont {Terning}}]{Bellazzini:2013fga}%
  \BibitemOpen
  \bibfield  {author} {\bibinfo {author} {\bibfnamefont {B.}~\bibnamefont
  {Bellazzini}}, \bibinfo {author} {\bibfnamefont {C.}~\bibnamefont {Csaki}},
  \bibinfo {author} {\bibfnamefont {J.}~\bibnamefont {Hubisz}}, \bibinfo
  {author} {\bibfnamefont {J.}~\bibnamefont {Serra}},\ and\ \bibinfo {author}
  {\bibfnamefont {J.}~\bibnamefont {Terning}},\ }\href
  {https://doi.org/10.1140/epjc/s10052-014-2790-x} {\bibfield  {journal}
  {\bibinfo  {journal} {Eur. Phys. J. C}\ }\textbf {\bibinfo {volume} {74}},\
  \bibinfo {pages} {2790} (\bibinfo {year} {2014})},\ \Eprint
  {https://arxiv.org/abs/1305.3919} {arXiv:1305.3919 [hep-th]} \BibitemShut
  {NoStop}%
\bibitem [{\citenamefont {Chacko}\ and\ \citenamefont
  {Mishra}(2013)}]{Chacko:2012sy}%
  \BibitemOpen
  \bibfield  {author} {\bibinfo {author} {\bibfnamefont {Z.}~\bibnamefont
  {Chacko}}\ and\ \bibinfo {author} {\bibfnamefont {R.~K.}\ \bibnamefont
  {Mishra}},\ }\href {https://doi.org/10.1103/PhysRevD.87.115006} {\bibfield
  {journal} {\bibinfo  {journal} {Phys. Rev. D}\ }\textbf {\bibinfo {volume}
  {87}},\ \bibinfo {pages} {115006} (\bibinfo {year} {2013})},\ \Eprint
  {https://arxiv.org/abs/1209.3022} {arXiv:1209.3022 [hep-ph]} \BibitemShut
  {NoStop}%
\bibitem [{\citenamefont {Holthausen}\ \emph {et~al.}(2013)\citenamefont
  {Holthausen}, \citenamefont {Kubo}, \citenamefont {Lim},\ and\ \citenamefont
  {Lindner}}]{Holthausen:2013ota}%
  \BibitemOpen
  \bibfield  {author} {\bibinfo {author} {\bibfnamefont {M.}~\bibnamefont
  {Holthausen}}, \bibinfo {author} {\bibfnamefont {J.}~\bibnamefont {Kubo}},
  \bibinfo {author} {\bibfnamefont {K.~S.}\ \bibnamefont {Lim}},\ and\ \bibinfo
  {author} {\bibfnamefont {M.}~\bibnamefont {Lindner}},\ }\href
  {https://doi.org/10.1007/JHEP12(2013)076} {\bibfield  {journal} {\bibinfo
  {journal} {JHEP}\ }\textbf {\bibinfo {volume} {12}},\ \bibinfo {pages}
  {076}},\ \Eprint {https://arxiv.org/abs/1310.4423} {arXiv:1310.4423 [hep-ph]}
  \BibitemShut {NoStop}%
\bibitem [{\citenamefont {Aguilar}\ \emph {et~al.}(2016)\citenamefont {Aguilar}
  \emph {et~al.}}]{AMS:2016oqu}%
  \BibitemOpen
  \bibfield  {author} {\bibinfo {author} {\bibfnamefont {M.}~\bibnamefont
  {Aguilar}} \emph {et~al.} (\bibinfo {collaboration} {AMS}),\ }\href
  {https://doi.org/10.1103/PhysRevLett.117.091103} {\bibfield  {journal}
  {\bibinfo  {journal} {Phys. Rev. Lett.}\ }\textbf {\bibinfo {volume} {117}},\
  \bibinfo {pages} {091103} (\bibinfo {year} {2016})}\BibitemShut {NoStop}%
\bibitem [{\citenamefont {Wang}\ and\ \citenamefont
  {Duan}(2025)}]{Wang:2025zth}%
  \BibitemOpen
  \bibfield  {author} {\bibinfo {author} {\bibfnamefont {X.}~\bibnamefont
  {Wang}}\ and\ \bibinfo {author} {\bibfnamefont {K.-K.}\ \bibnamefont
  {Duan}},\ }\href@noop {} {\  (\bibinfo {year} {2025})},\ \Eprint
  {https://arxiv.org/abs/2512.12176} {arXiv:2512.12176 [astro-ph.HE]}
  \BibitemShut {NoStop}%
\bibitem [{\citenamefont {Cassel}(2010)}]{Cassel:2009wt}%
  \BibitemOpen
  \bibfield  {author} {\bibinfo {author} {\bibfnamefont {S.}~\bibnamefont
  {Cassel}},\ }\href {https://doi.org/10.1088/0954-3899/37/10/105009}
  {\bibfield  {journal} {\bibinfo  {journal} {J. Phys. G}\ }\textbf {\bibinfo
  {volume} {37}},\ \bibinfo {pages} {105009} (\bibinfo {year} {2010})},\
  \Eprint {https://arxiv.org/abs/0903.5307} {arXiv:0903.5307 [hep-ph]}
  \BibitemShut {NoStop}%
\bibitem [{\citenamefont {Slatyer}(2010)}]{Slatyer:2009vg}%
  \BibitemOpen
  \bibfield  {author} {\bibinfo {author} {\bibfnamefont {T.~R.}\ \bibnamefont
  {Slatyer}},\ }\href {https://doi.org/10.1088/1475-7516/2010/02/028}
  {\bibfield  {journal} {\bibinfo  {journal} {JCAP}\ }\textbf {\bibinfo
  {volume} {02}},\ \bibinfo {pages} {028}},\ \Eprint
  {https://arxiv.org/abs/0910.5713} {arXiv:0910.5713 [hep-ph]} \BibitemShut
  {NoStop}%
\bibitem [{\citenamefont {McDaniel}\ \emph {et~al.}(2024)\citenamefont
  {McDaniel}, \citenamefont {Ajello}, \citenamefont {Karwin}, \citenamefont
  {Di~Mauro}, \citenamefont {Drlica-Wagner},\ and\ \citenamefont
  {S{\'a}nchez-Conde}}]{McDaniel:2023bju}%
  \BibitemOpen
  \bibfield  {author} {\bibinfo {author} {\bibfnamefont {A.}~\bibnamefont
  {McDaniel}}, \bibinfo {author} {\bibfnamefont {M.}~\bibnamefont {Ajello}},
  \bibinfo {author} {\bibfnamefont {C.~M.}\ \bibnamefont {Karwin}}, \bibinfo
  {author} {\bibfnamefont {M.}~\bibnamefont {Di~Mauro}}, \bibinfo {author}
  {\bibfnamefont {A.}~\bibnamefont {Drlica-Wagner}},\ and\ \bibinfo {author}
  {\bibfnamefont {M.~A.}\ \bibnamefont {S{\'a}nchez-Conde}},\ }\href
  {https://doi.org/10.1103/PhysRevD.109.063024} {\bibfield  {journal} {\bibinfo
   {journal} {Phys. Rev. D}\ }\textbf {\bibinfo {volume} {109}},\ \bibinfo
  {pages} {063024} (\bibinfo {year} {2024})},\ \Eprint
  {https://arxiv.org/abs/2311.04982} {arXiv:2311.04982 [astro-ph.HE]}
  \BibitemShut {NoStop}%
\bibitem [{\citenamefont {Cirelli}\ \emph {et~al.}(2011)\citenamefont
  {Cirelli}, \citenamefont {Corcella}, \citenamefont {Hektor}, \citenamefont
  {Hutsi}, \citenamefont {Kadastik}, \citenamefont {Panci}, \citenamefont
  {Raidal}, \citenamefont {Sala},\ and\ \citenamefont
  {Strumia}}]{Cirelli:2010xx}%
  \BibitemOpen
  \bibfield  {author} {\bibinfo {author} {\bibfnamefont {M.}~\bibnamefont
  {Cirelli}}, \bibinfo {author} {\bibfnamefont {G.}~\bibnamefont {Corcella}},
  \bibinfo {author} {\bibfnamefont {A.}~\bibnamefont {Hektor}}, \bibinfo
  {author} {\bibfnamefont {G.}~\bibnamefont {Hutsi}}, \bibinfo {author}
  {\bibfnamefont {M.}~\bibnamefont {Kadastik}}, \bibinfo {author}
  {\bibfnamefont {P.}~\bibnamefont {Panci}}, \bibinfo {author} {\bibfnamefont
  {M.}~\bibnamefont {Raidal}}, \bibinfo {author} {\bibfnamefont
  {F.}~\bibnamefont {Sala}},\ and\ \bibinfo {author} {\bibfnamefont
  {A.}~\bibnamefont {Strumia}},\ }\href
  {https://doi.org/10.1088/1475-7516/2012/10/E01} {\bibfield  {journal}
  {\bibinfo  {journal} {JCAP}\ }\textbf {\bibinfo {volume} {03}},\ \bibinfo
  {pages} {051}},\ \bibinfo {note} {[Erratum: JCAP 10, E01 (2012)]},\ \Eprint
  {https://arxiv.org/abs/1012.4515} {arXiv:1012.4515 [hep-ph]} \BibitemShut
  {NoStop}%
\bibitem [{\citenamefont {Boddy}\ \emph {et~al.}(2016)\citenamefont {Boddy},
  \citenamefont {Dienes}, \citenamefont {Kim}, \citenamefont {Kumar},
  \citenamefont {Park},\ and\ \citenamefont {Thomas}}]{Boddy:2016fds}%
  \BibitemOpen
  \bibfield  {author} {\bibinfo {author} {\bibfnamefont {K.~K.}\ \bibnamefont
  {Boddy}}, \bibinfo {author} {\bibfnamefont {K.~R.}\ \bibnamefont {Dienes}},
  \bibinfo {author} {\bibfnamefont {D.}~\bibnamefont {Kim}}, \bibinfo {author}
  {\bibfnamefont {J.}~\bibnamefont {Kumar}}, \bibinfo {author} {\bibfnamefont
  {J.-C.}\ \bibnamefont {Park}},\ and\ \bibinfo {author} {\bibfnamefont
  {B.}~\bibnamefont {Thomas}},\ }\href
  {https://doi.org/10.1103/PhysRevD.94.095027} {\bibfield  {journal} {\bibinfo
  {journal} {Phys. Rev. D}\ }\textbf {\bibinfo {volume} {94}},\ \bibinfo
  {pages} {095027} (\bibinfo {year} {2016})},\ \Eprint
  {https://arxiv.org/abs/1606.07440} {arXiv:1606.07440 [hep-ph]} \BibitemShut
  {NoStop}%
\bibitem [{\citenamefont {Winkler}(2019)}]{Winkler:2018qyg}%
  \BibitemOpen
  \bibfield  {author} {\bibinfo {author} {\bibfnamefont {M.~W.}\ \bibnamefont
  {Winkler}},\ }\href {https://doi.org/10.1103/PhysRevD.99.015018} {\bibfield
  {journal} {\bibinfo  {journal} {Phys. Rev. D}\ }\textbf {\bibinfo {volume}
  {99}},\ \bibinfo {pages} {015018} (\bibinfo {year} {2019})},\ \Eprint
  {https://arxiv.org/abs/1809.01876} {arXiv:1809.01876 [hep-ph]} \BibitemShut
  {NoStop}%
\bibitem [{\citenamefont {Aalbers}\ \emph {et~al.}(2025)\citenamefont {Aalbers}
  \emph {et~al.}}]{LZ:2024zvo}%
  \BibitemOpen
  \bibfield  {author} {\bibinfo {author} {\bibfnamefont {J.}~\bibnamefont
  {Aalbers}} \emph {et~al.} (\bibinfo {collaboration} {LZ}),\ }\href
  {https://doi.org/10.1103/4dyc-z8zf} {\bibfield  {journal} {\bibinfo
  {journal} {Phys. Rev. Lett.}\ }\textbf {\bibinfo {volume} {135}},\ \bibinfo
  {pages} {011802} (\bibinfo {year} {2025})},\ \Eprint
  {https://arxiv.org/abs/2410.17036} {arXiv:2410.17036 [hep-ex]} \BibitemShut
  {NoStop}%
\bibitem [{\citenamefont {Aprile}\ \emph {et~al.}(2025)\citenamefont {Aprile}
  \emph {et~al.}}]{XENON:2025vwd}%
  \BibitemOpen
  \bibfield  {author} {\bibinfo {author} {\bibfnamefont {E.}~\bibnamefont
  {Aprile}} \emph {et~al.} (\bibinfo {collaboration} {XENON}),\ }\href
  {https://doi.org/10.1103/msw4-t342} {\bibfield  {journal} {\bibinfo
  {journal} {Phys. Rev. Lett.}\ }\textbf {\bibinfo {volume} {135}},\ \bibinfo
  {pages} {221003} (\bibinfo {year} {2025})},\ \Eprint
  {https://arxiv.org/abs/2502.18005} {arXiv:2502.18005 [hep-ex]} \BibitemShut
  {NoStop}%
\bibitem [{\citenamefont {Haisch}\ and\ \citenamefont
  {Kamenik}(2016)}]{Haisch:2016hzu}%
  \BibitemOpen
  \bibfield  {author} {\bibinfo {author} {\bibfnamefont {U.}~\bibnamefont
  {Haisch}}\ and\ \bibinfo {author} {\bibfnamefont {J.~F.}\ \bibnamefont
  {Kamenik}},\ }\href {https://doi.org/10.1103/PhysRevD.93.055047} {\bibfield
  {journal} {\bibinfo  {journal} {Phys. Rev. D}\ }\textbf {\bibinfo {volume}
  {93}},\ \bibinfo {pages} {055047} (\bibinfo {year} {2016})},\ \Eprint
  {https://arxiv.org/abs/1601.05110} {arXiv:1601.05110 [hep-ph]} \BibitemShut
  {NoStop}%
\bibitem [{\citenamefont {Berdugo}(2022)}]{Berdugo:2022zzo}%
  \BibitemOpen
  \bibfield  {author} {\bibinfo {author} {\bibfnamefont {J.}~\bibnamefont
  {Berdugo}} (\bibinfo {collaboration} {AMS}),\ }\href
  {https://doi.org/10.3103/S0027134922020126} {\bibfield  {journal} {\bibinfo
  {journal} {Moscow Univ. Phys. Bull.}\ }\textbf {\bibinfo {volume} {77}},\
  \bibinfo {pages} {71} (\bibinfo {year} {2022})}\BibitemShut {NoStop}%
\bibitem [{\citenamefont {Alimena}\ \emph {et~al.}(2020)\citenamefont {Alimena}
  \emph {et~al.}}]{Alimena:2019zri}%
  \BibitemOpen
  \bibfield  {author} {\bibinfo {author} {\bibfnamefont {J.}~\bibnamefont
  {Alimena}} \emph {et~al.},\ }\href {https://doi.org/10.1088/1361-6471/ab4574}
  {\bibfield  {journal} {\bibinfo  {journal} {J. Phys. G}\ }\textbf {\bibinfo
  {volume} {47}},\ \bibinfo {pages} {090501} (\bibinfo {year} {2020})},\
  \Eprint {https://arxiv.org/abs/1903.04497} {arXiv:1903.04497 [hep-ex]}
  \BibitemShut {NoStop}%
\bibitem [{\citenamefont {Alekhin}\ \emph {et~al.}(2016)\citenamefont {Alekhin}
  \emph {et~al.}}]{Alekhin:2015byh}%
  \BibitemOpen
  \bibfield  {author} {\bibinfo {author} {\bibfnamefont {S.}~\bibnamefont
  {Alekhin}} \emph {et~al.},\ }\href
  {https://doi.org/10.1088/0034-4885/79/12/124201} {\bibfield  {journal}
  {\bibinfo  {journal} {Rept. Prog. Phys.}\ }\textbf {\bibinfo {volume} {79}},\
  \bibinfo {pages} {124201} (\bibinfo {year} {2016})},\ \Eprint
  {https://arxiv.org/abs/1504.04855} {arXiv:1504.04855 [hep-ph]} \BibitemShut
  {NoStop}%
\bibitem [{\citenamefont {Curtin}\ \emph {et~al.}(2019)\citenamefont {Curtin}
  \emph {et~al.}}]{Curtin:2018mvb}%
  \BibitemOpen
  \bibfield  {author} {\bibinfo {author} {\bibfnamefont {D.}~\bibnamefont
  {Curtin}} \emph {et~al.},\ }\href {https://doi.org/10.1088/1361-6633/ab28d6}
  {\bibfield  {journal} {\bibinfo  {journal} {Rept. Prog. Phys.}\ }\textbf
  {\bibinfo {volume} {82}},\ \bibinfo {pages} {116201} (\bibinfo {year}
  {2019})},\ \Eprint {https://arxiv.org/abs/1806.07396} {arXiv:1806.07396
  [hep-ph]} \BibitemShut {NoStop}%
\bibitem [{\citenamefont {Flowers}\ \emph {et~al.}(2020)\citenamefont
  {Flowers}, \citenamefont {Meier}, \citenamefont {Rogan}, \citenamefont
  {Kang},\ and\ \citenamefont {Park}}]{Flowers:2019gvj}%
  \BibitemOpen
  \bibfield  {author} {\bibinfo {author} {\bibfnamefont {Z.}~\bibnamefont
  {Flowers}}, \bibinfo {author} {\bibfnamefont {Q.}~\bibnamefont {Meier}},
  \bibinfo {author} {\bibfnamefont {C.}~\bibnamefont {Rogan}}, \bibinfo
  {author} {\bibfnamefont {D.~W.}\ \bibnamefont {Kang}},\ and\ \bibinfo
  {author} {\bibfnamefont {S.~C.}\ \bibnamefont {Park}},\ }\href
  {https://doi.org/10.1007/JHEP03(2020)132} {\bibfield  {journal} {\bibinfo
  {journal} {JHEP}\ }\textbf {\bibinfo {volume} {03}},\ \bibinfo {pages}
  {132}},\ \Eprint {https://arxiv.org/abs/1903.05825} {arXiv:1903.05825
  [hep-ph]} \BibitemShut {NoStop}%
\bibitem [{\citenamefont {Ariga}\ \emph {et~al.}(2019)\citenamefont {Ariga}
  \emph {et~al.}}]{FASER:2019aik}%
  \BibitemOpen
  \bibfield  {author} {\bibinfo {author} {\bibfnamefont {A.}~\bibnamefont
  {Ariga}} \emph {et~al.} (\bibinfo {collaboration} {FASER}),\ }\href@noop {}
  {\  (\bibinfo {year} {2019})},\ \Eprint {https://arxiv.org/abs/1901.04468}
  {arXiv:1901.04468 [hep-ex]} \BibitemShut {NoStop}%
\bibitem [{\citenamefont {Abada}\ \emph {et~al.}(2019)\citenamefont {Abada}
  \emph {et~al.}}]{FCC:2018evy}%
  \BibitemOpen
  \bibfield  {author} {\bibinfo {author} {\bibfnamefont {A.}~\bibnamefont
  {Abada}} \emph {et~al.} (\bibinfo {collaboration} {FCC}),\ }\href
  {https://doi.org/10.1140/epjst/e2019-900045-4} {\bibfield  {journal}
  {\bibinfo  {journal} {Eur. Phys. J. ST}\ }\textbf {\bibinfo {volume} {228}},\
  \bibinfo {pages} {261} (\bibinfo {year} {2019})}\BibitemShut {NoStop}%
\bibitem [{\citenamefont {Blondel}\ \emph {et~al.}(2022)\citenamefont {Blondel}
  \emph {et~al.}}]{Blondel:2022qqo}%
  \BibitemOpen
  \bibfield  {author} {\bibinfo {author} {\bibfnamefont {A.}~\bibnamefont
  {Blondel}} \emph {et~al.},\ }\href {https://doi.org/10.3389/fphy.2022.967881}
  {\bibfield  {journal} {\bibinfo  {journal} {Front. in Phys.}\ }\textbf
  {\bibinfo {volume} {10}},\ \bibinfo {pages} {967881} (\bibinfo {year}
  {2022})},\ \Eprint {https://arxiv.org/abs/2203.05502} {arXiv:2203.05502
  [hep-ex]} \BibitemShut {NoStop}%
\end{thebibliography}%

\clearpage
\onecolumngrid
\begin{center}
   \textbf{\large APPENDIX: SUPPLEMENTARY MATERIAL \\[.2cm] ``Aspects of Sommerfeld Enhancement in the light of Halo gamma-ray excess''}\\[.2cm]
  \vspace{0.05in}
  {Yongsoo Jho, Jeonghwan Park, Min Gi Park, Seong Chan Park}
\end{center}

\setcounter{equation}{0}
\setcounter{figure}{0}
\setcounter{table}{0}
\setcounter{section}{0}
\setcounter{page}{1}
\makeatletter
\renewcommand{\theequation}{S\arabic{equation}}
\renewcommand{\thefigure}{S\arabic{figure}}
\renewcommand{\thetable}{S\arabic{table}}

\onecolumngrid

\section{Heuristic understanding of Sommerfeld enhancement}

\begin{figure}[h]
\includegraphics[width=0.54\textwidth]{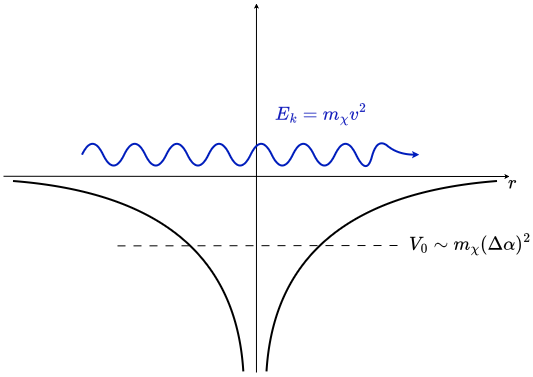}
\caption{A schematic representation of the resonant scattering taking place when the kinetic energy ($E_k$) is matching to the binding energy ($E_b = V_0$).} 
\label{fig:schematic_scattering}
\end{figure}

The saturation occurs due to the resonance at $E_k \simeq E_b$ or
\begin{align}
      m_\chi v_{\rm sat}^2 \sim m_\chi (\Delta \alpha)^2 \Rightarrow 
      v_{\rm sat}\sim \Delta \alpha,
\end{align}
where we took the energy scales of the scattering:
\begin{itemize}
    \item \textbf{Kinetic Energy ($E_k$)}: Incoming energy of the dark matter.
    \begin{align}
        E_k = \frac{1}{2}\mu v_{rel}^2 = m_\chi v^2.
    \end{align}
    \item \textbf{Binding Energy ($E_b$)}: Depth of the shallow bound state. Since the scattering length scales as $a \sim [m_\chi (\alpha - \alpha_{\rm crit})]^{-1}$, the binding energy is:
    \begin{align}
        E_b = \frac{1}{2\mu a^2} \sim m_\chi (\Delta \alpha)^2.
    \end{align}
\end{itemize}

\section{Full Numerical Solution of the Schr\"odinger equation}
The Schr\"odinger equation for the scattering wavefunction in a Yukawa potential,
$V(r)=-\alpha\frac{e^{-m_\varphi r}}{r}$
is
\begin{align}
\left(-\frac{\nabla^2}{2\mu}-\alpha\frac{e^{-m_\varphi r}}{r}-\frac{k^2}{2\mu} - i \frac{\Gamma(\vec{r})}{2} \right)\chi_{\vec{k}}(\vec r)=0,
\label{eq:schro_3d}
\end{align}
where $\mu=m_\chi/2$ is the reduced mass. In the non-relativistic limit we take
$k=m_\chi v$, with $v$ being the speed of each particle in the center-of-mass frame. The decay term $\Gamma (\vec{r}) \sim \frac{\Gamma}{V} \Theta (r_{\rm cut} -r)$, with a scale of the hard process $r_{\rm cut} \sim m_\chi^{-1}$, becomes important when $\Gamma/m_\chi \gg \Delta \alpha$. In our scenario, $\Gamma \simeq \frac{3}{192} \alpha^7 m_\chi$ and the off-resonance $\Delta \alpha$ dominantly determine the saturation velocity since $\Delta \alpha \gg (\Gamma/m_\chi)^{1/2}$ which is satisfied in our parameter choice. Thus, we ignore the decay term, and focus on the effect by an off-resonance $\Delta \alpha$ directly obtaining a solution of the equation.

For a spherically symmetric potential, we expand
\begin{align}
\chi_{\vec{k}}(\vec r)=\sum_{l,m} R_{l,k}(r)\,Y_{lm}(\Omega),
\end{align}
and introduce the reduced radial wavefunction
\begin{align}
u_{l,k}(r)\equiv r\,R_{l,k}(r).
\end{align}
Then the radial equation becomes
\begin{align}
u_{l,k}''(r)+\left[k^2-\frac{l(l+1)}{r^2}+m_\chi\alpha\,\frac{e^{-m_\varphi r}}{r}\right]u_{l,k}(r)=0.
\label{eq:radial_r}
\end{align}

It is convenient to introduce dimensionless variables
\begin{align}
x\equiv kr,\qquad \epsilon\equiv \frac{m_\varphi}{k},
\end{align}
and define
\begin{align}
w_{l,k}(x)\equiv k\,u_{l,k}\!\left(\frac{x}{k}\right).
\end{align}
With these, Eq.~\eqref{eq:radial_r} can be rewritten as
\begin{align}
w_{l,k}''(x)+\left[\,1-\frac{l(l+1)}{x^2}+\frac{\alpha}{v}\frac{e^{-\epsilon x}}{x}\right]w_{l,k}(x)=0,
\label{eq:radial_x}
\end{align}
where primes denote derivatives with respect to $x$.
\paragraph{\textbf{Large-$x$ limit ($x\to\infty$).}}
In the large-$x$ regime the Yukawa interaction is exponentially suppressed,
$e^{-\epsilon x}/x\to 0$, and the centrifugal term is also negligible for $x\gg 1$.
Thus Eq.~\eqref{eq:radial_x} reduces to
\begin{align}
w_{l,k}''(x)+w_{l,k}(x)\simeq 0
\qquad (x\to\infty),
\label{eq:far_eq_simple}
\end{align}
with the asymptotic solution
\begin{align}
w_{l,k}(x)\simeq C_{l,k}\sin(x+\delta_l)
\label{eq:far_sol_simple}
\end{align}

\paragraph{\textbf{Small-$x$ limit ($x\to 0$).}}
In the small-$x$ regime, the angular-momentum barrier dominates the dynamics, so
Eq.~\eqref{eq:radial_x} simplifies to
\begin{align}
w_{l,k}''(x)-\frac{l(l+1)}{x^2}w_{l,k}(x)\simeq 0
\qquad (x\to 0).
\label{eq:near_eq_simple}
\end{align}
The general solution is
\begin{align}
w_{l,k}(x)\simeq A_{l,k}x^{l+1}+B_{l,k}x^{-l},
\label{eq:near_sol_simple}
\end{align}
and regularity at the origin selects the physical branch
\begin{align}
w_{l,k}(x)\simeq A_{l,k}x^{l+1}.
\label{eq:near_sol_reg}
\end{align}

\paragraph{\textbf{Normalization and Sommerfeld enhancement.}}
The numerical solution obtained from Eq.~\eqref{eq:radial_x} is defined up to an overall
constant. In the far-field limit we write
\begin{align}
w_{l,k}(x)\xrightarrow[x\to\infty]{} C_{l,k}\,\sin(x+\delta_l),
\label{eq:far_C}
\end{align}
while regularity at the origin implies the near-field behaviour
\begin{align}
w_{l,k}(x)\xrightarrow[x\to 0]{} A_{l,k}\,x^{l+1}.
\label{eq:near_A}
\end{align}
The continuum normalization condition~\cite{Iengo:2009ni},
\begin{align}
\int_0^\infty dr\,u_{l,q}(r)\,u_{l,k}(r)=\delta(k-q),
\label{eq:cont_norm}
\end{align}
fixes the asymptotic amplitude of $u_{l,k}(r)$ as
\begin{align}
u_{l,k}(r)\xrightarrow[r\to\infty]{}\sqrt{\frac{2}{\pi}}\sin(kr+\delta_l),
\label{eq:u_asym_norm}
\end{align}
and therefore, for $w_{l,k}(x)\equiv k\,u_{l,k}(x/k)$ with $x\equiv kr$,
\begin{align}
w_{l,k}^{\rm (norm)}(x)\xrightarrow[x\to\infty]{}k\sqrt{\frac{2}{\pi}}\sin(x+\delta_l).
\label{eq:w_asym_norm}
\end{align}
Comparing Eq.~\eqref{eq:far_C} with Eq.~\eqref{eq:w_asym_norm}, the normalized solution is
\begin{align}
w_{l,k}^{\rm (norm)}(x)=\frac{k\sqrt{2/\pi}}{C_{l,k}}\,w_{l,k}(x).
\label{eq:w_norm_rescale}
\end{align}
Consequently, its near-field coefficient becomes
\begin{align}
A_{l,k}^{\rm (norm)}=\frac{k\sqrt{2/\pi}}{C_{l,k}}\,A_{l,k}.
\label{eq:A_norm}
\end{align}

\begin{figure*}[t]
\begin{minipage}[t]{0.55\linewidth}
    \centering
    \includegraphics[width=\linewidth]{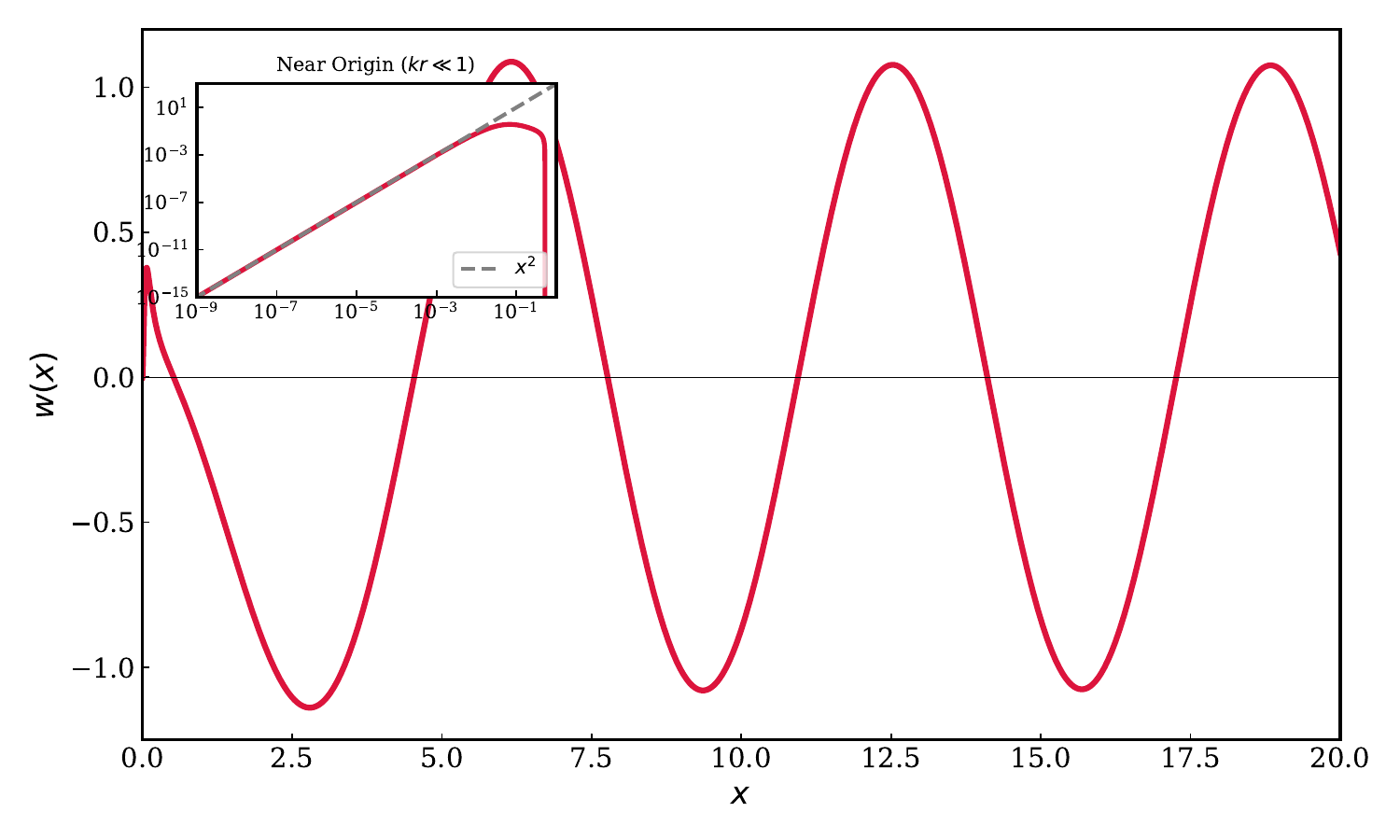}
    \caption{Numerical solution of the radial wavefunction $w(x)$ as a function of $x\equiv kr$ for Yukawa scattering in the $l=1$ (p-wave) channel, using the same parameter set as Fig.~\ref{fig:xsec_velocity_dep} with $v=10^{-3}$. The inset shows the near-origin behavior, confirming the expected regular scaling $w(x)\propto x^{2}$ (grey dashed line) for $x\ll 1$.}
    \label{fig:wavefunction}
\end{minipage}\hfill
\begin{minipage}[t]{0.44\linewidth}
    \centering
    \includegraphics[width=\linewidth]{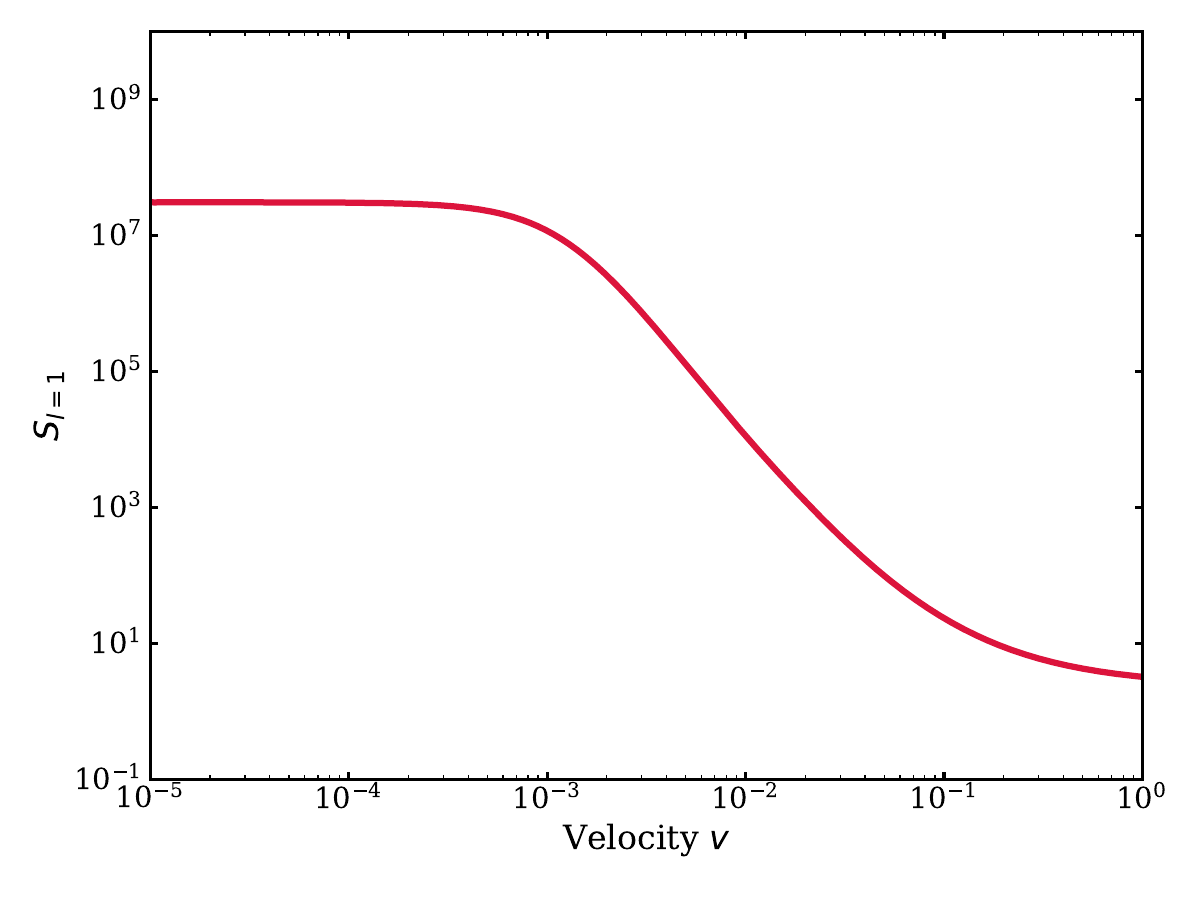}
    \caption{Velocity dependence of the Sommerfeld enhancement factor in the $l=1$ (p-wave) channel, $S_{l=1}(v)$, for Yukawa interactions with the same parameter set as Fig.~\ref{fig:xsec_velocity_dep}. The enhancement saturates at low velocities due to the finite range of the mediator. Also, the resulting cross section decreases toward the perturbative limit at larger $v$.}
    \label{fig:Sommerfeldfactor}
\end{minipage}
\end{figure*}

Using $u_{l,k}=w_{l,k}/k$ and $R_{l,k}=u_{l,k}/r$, the normalized radial wavefunction near
the origin behaves as
\begin{align}
R_{l,k}^{\rm (norm)}(r)
=\frac{u_{l,k}^{\rm (norm)}(r)}{r}
\simeq A_{l,k}^{\rm (norm)}\,k^{l}\,r^{\,l}
\qquad (r\to 0),
\label{eq:R_near}
\end{align}
so that
\begin{align}
\left.\frac{\partial^{l}R_{l,k}^{\rm (norm)}(r)}{\partial r^{l}}\right|_{r=0}
= l!\,A_{l,k}^{\rm (norm)}\,k^{l}.
\label{eq:R_deriv0}
\end{align}
Substituting Eq.~\eqref{eq:R_deriv0} into
\begin{align}
S_l=\left|\left.\frac{(2l+1)!!}{k^{l}\,l!}\,
\frac{\partial^{l}R_{l,k}(r)}{\partial r^{l}}\right|_{r=0}\right|^2,
\label{eq:Sl_def}
\end{align}
we obtain
\begin{align}
S_l=\left|(2l+1)!!\,A_{l,k}^{\rm (norm)}\right|^2
=\left|(2l+1)!!\,\frac{k\sqrt{2/\pi}}{C_{l,k}}\,A_{l,k}\right|^2.
\label{eq:Sl_final}
\end{align}
In practice, one may solve Eq.~\eqref{eq:radial_x} with arbitrary overall normalization,
extract $A_{l,k}$ from the $x\to 0$ behavior and $C_{l,k}$ from the $x\to\infty$ behavior,
and then evaluate $S_l$ via Eq.~\eqref{eq:Sl_final}.
Fig.~\ref{fig:wavefunction} shows the numerical solution for the wavefunction $w(x)$ using the same parameter values as in Fig.~\ref{fig:xsec_velocity_dep}.
\section{Lorentz boost effect in the gamma-ray spectrum}

\subsection{Kinematics \& Motivation}
We consider the regime where the dark matter mass is significantly heavier than the mediator mass ($m_\chi \gg m_\varphi$). In this limit, the mediators produced from the annihilation process $\chi\chi \to \varphi\varphi$ are highly boosted. For our benchmark parameters ($m_\chi=575$ GeV and $m_\varphi=11.6$ GeV), the Lorentz factor reaches $\gamma = m_\chi/m_\varphi \approx 49.57$. This large boost leads to a significant broadening of the gamma-ray spectrum in the laboratory frame, extending the maximum energy up to $\sim m_\chi$ while flattening the low-energy distribution.

\subsection{Analytical Formula}
Let $E'$ be the photon energy in the $\varphi$ rest frame. In the laboratory frame, the photon energy $E$ is given by the standard Doppler shift formula:
\begin{align}
    E = \gamma E'( 1 +\beta \cos \theta'),
\end{align}
where $\beta = \sqrt{1-\gamma^{-2}}$ and $\theta'$ is the emission angle in the rest frame. Since $\varphi$ is a scalar, it decays isotropically, meaning the angular distribution is uniform: $dN/d(\cos \theta') = 1/2$ for $\cos\theta' \in [-1, 1]$.

We derive the lab-frame spectrum $dN/dE$ using the Jacobian of the transformation. From the relation $dE = \gamma \beta E' d(\cos \theta')$, the distribution becomes:
\begin{align}
    \frac{dN}{dE} = \frac{dN}{d(\cos \theta')} \left| \frac{d (\cos \theta')}{d E} \right| = \frac{1}{2\gamma \beta E'}.
\end{align}
For a monochromatic line at $E'$, this results in a flat "box" spectrum spanning the range $E_- \leq E \leq E_+$, where the kinematic bounds are $E_{\pm} = \gamma E'(1 \pm \beta)$.

The final gamma-ray spectrum is obtained by convolving the rest-frame source spectrum, $(dN/dE')_{\text{rest}}$, with this box kernel:
\begin{align}
    \frac{dN_\text{lab}}{dE} = \int_{E'_{min}}^{E'_{max}} \left(\frac{dN}{dE'}\right)_{\text{rest}} \frac{1}{2\gamma \beta E'} dE'.
\end{align}
Here, the integration limits are set by the kinematic constraints:
\begin{align}
    E'_{min} = \frac{E}{\gamma(1+\beta)}, \quad E'_{max} = \min\left[\frac{E}{\gamma(1-\beta)}, \frac{m_\varphi}{2}\right].
\end{align}

\subsection{Numerical Implementation and Results}
\begin{figure}[h]
\includegraphics[width=0.5\textwidth]{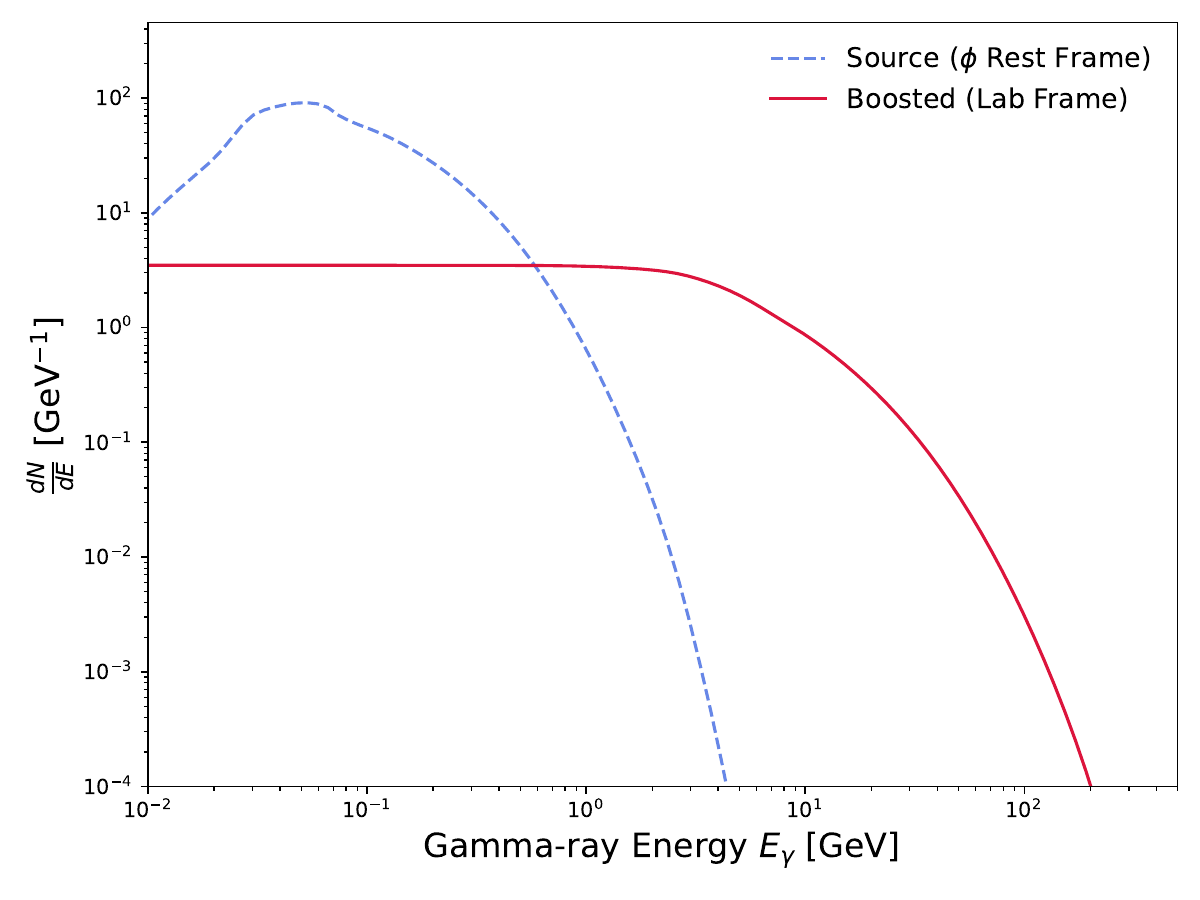}
\caption{The spectrum with and without including the Lorentz boost effect. The dashed blue line represents the spectrum in the rest frame of the mediator ($m_\varphi = 11.6$ GeV). The solid red line shows the boosted spectrum in the laboratory frame, assuming dark matter mass $m_\chi = 575$ GeV, corresponding to a Lorentz boost factor of $\gamma = m_\chi/m_\varphi = 49.57$. The boosting effect significantly broadens the spectrum up to $E_\gamma \approx m_\chi$ and flattens the distribution at lower energies.}
    \label{fig:boost_effect}
\end{figure}

To illustrate the spectral broadening, we perform a numerical evaluation using the parameters $m_\chi = 575$ GeV and $m_\varphi = 11.6$ GeV. Fig.~\ref{fig:boost_effect} presents the resulting gamma-ray spectra.

The key features of our numerical results are summarized as follows:

\begin{itemize}
    \item \textbf{Source Spectrum}: \\
    We obtain the rest-frame source spectrum, $(dN/dE')_\text{rest}$, from the PPPC4DMID tables \cite{Cirelli:2010xx}. We assume the dominant decay channel is $\varphi \to b\bar{b}$ and set the parent particle mass to $m_\text{parent} = m_\varphi$.

    \item \textbf{Result}: \\
    The Lorentz boost modifies the spectral shape significantly:
    
    \paragraph{\textbf{Low Energy Regime}} The spectrum exhibits a flattened profile in the low-energy region. This flattening occurs because the boost effectively smears the low-energy photons over a much broader energy range.
    
    \paragraph{\textbf{High Energy Regime}} The spectrum undergoes a significant blue shift. The kinematic edge extends from the rest-frame limit of $m_\varphi/2$ up to approximately $\gamma m_\varphi (1+\beta) \approx m_\chi/2$ in the laboratory frame.
    
    \item \textbf{Comparison}: \\ As shown in Fig.~\ref{fig:boost_effect}, comparing the rest frame (dashed) and laboratory frame (solid) results confirms that while the total number of photons is conserved, the energy distribution is drastically reshaped. The narrow peak in the rest frame transforms into a broad box-like feature in the laboratory frame.
\end{itemize}

\end{document}